\documentclass{article}
\usepackage{hyperref}
\usepackage{astrobib}
\usepackage{epsfig}
\def \beginfig          {\begin{figure}}
\def \endfig            {\end{figure}}

\def \begineq           {\begin{equation}}
\def \endeq             {\end{equation}}

\def\gtorder{\mathrel{\raise.3ex\hbox{$>$}\mkern-14mu
             \lower0.6ex\hbox{$\sim$}}}
\def\ltorder{\mathrel{\raise.3ex\hbox{$<$}\mkern-14mu
             \lower0.6ex\hbox{$\sim$}}}

\def \lsim {\ltorder}
\def \gsim {\gtorder}
\def \hide#1{}
\def \half {{1\over 2}}

\def\bk {{\bf k}}

\def\bm {{\bf m}}
\def\bn {{\bf n}}

\def\br {{\bf r}}

\def\bv {{\bf v}}

\def\bx {{\bf x}}

\def\bz {{\bf z}}

\def\bF {{\bf F}}

\def\bW {{\bf W}}

\def\btheta {\mbox{\boldmath $\theta$}}

\def\bkappa {\mbox{\boldmath $\kappa$}}

\def\bsigma {\mbox{\boldmath $\sigma$}}

\def\vtheta {{\vec \theta}}

\def\araa{{ARA\&A}}             
\def\apj{{ApJ}}                 
\def\apjs{{ApJS}}               
\def\aap{{A\&A}}                
\def\mnras{{MNRAS}}             
\def\pasp{{PASP}}               

\def\procspie{{Proc.~SPIE}}   

\def \figwidth {0.9 \linewidth}

\def \Dval	{1.5}
\def \Ntval	{36}
\def \Ntauval	{32}

\def \totalcost	{\$70{\rm M}}
\def \pixelsize		{0''.1}
\def \pixsizeinmicrons	{5}
\def \cellsizeinpix	{600}
\def \chipsizeinpix	{18{\rm K}}

\def \chipsizeincells	{30}

\def \FWHM		{0''.12}

\begin{document}

\title{A New Strategy for Deep Wide-Field High Resolution Optical Imaging.}
\author{N.~Kaiser, J.L.~Tonry and G.A.~Luppino \\
Institute for Astronomy, U.~Hawaii}
\maketitle

\begin{abstract}{
We propose a new strategy for obtaining enhanced resolution (FWHM $\simeq \FWHM$)
deep optical images over a wide field of view.  As is well known, this type of image quality 
can be obtained in principle simply by fast guiding on a small ($D \sim 1.5$m) 
telescope at a good site, but only for target objects which lie within a limited angular
distance of a suitably bright guide star.  For high altitude turbulence this
`isokinetic angle' is approximately $1'$.
With a 1 degree field say one would need to track and correct
the motions of thousands of isokinetic patches, yet there are typically
too few sufficiently bright guide stars to provide the necessary guiding information.
Our proposed solution to these problems has two novel features.
The first is to use orthogonal transfer charge-coupled device
(OTCCD) technology to effectively implement a wide field `rubber focal plane'
detector composed of an array of cells which can be guided independently.  
The second is to combine measured motions of a set of guide stars made with
an array of telescopes to provide the extra information needed to
fully determine the deflection field. We discuss the
performance, feasibility and design constraints on a system which would
provide the collecting area equivalent to a single $9$m telescope, a
1 degree square field and $\simeq \FWHM$ FWHM image quality.
}\end{abstract}

\begin{figure}[htbp!]
\centering\epsfig{file=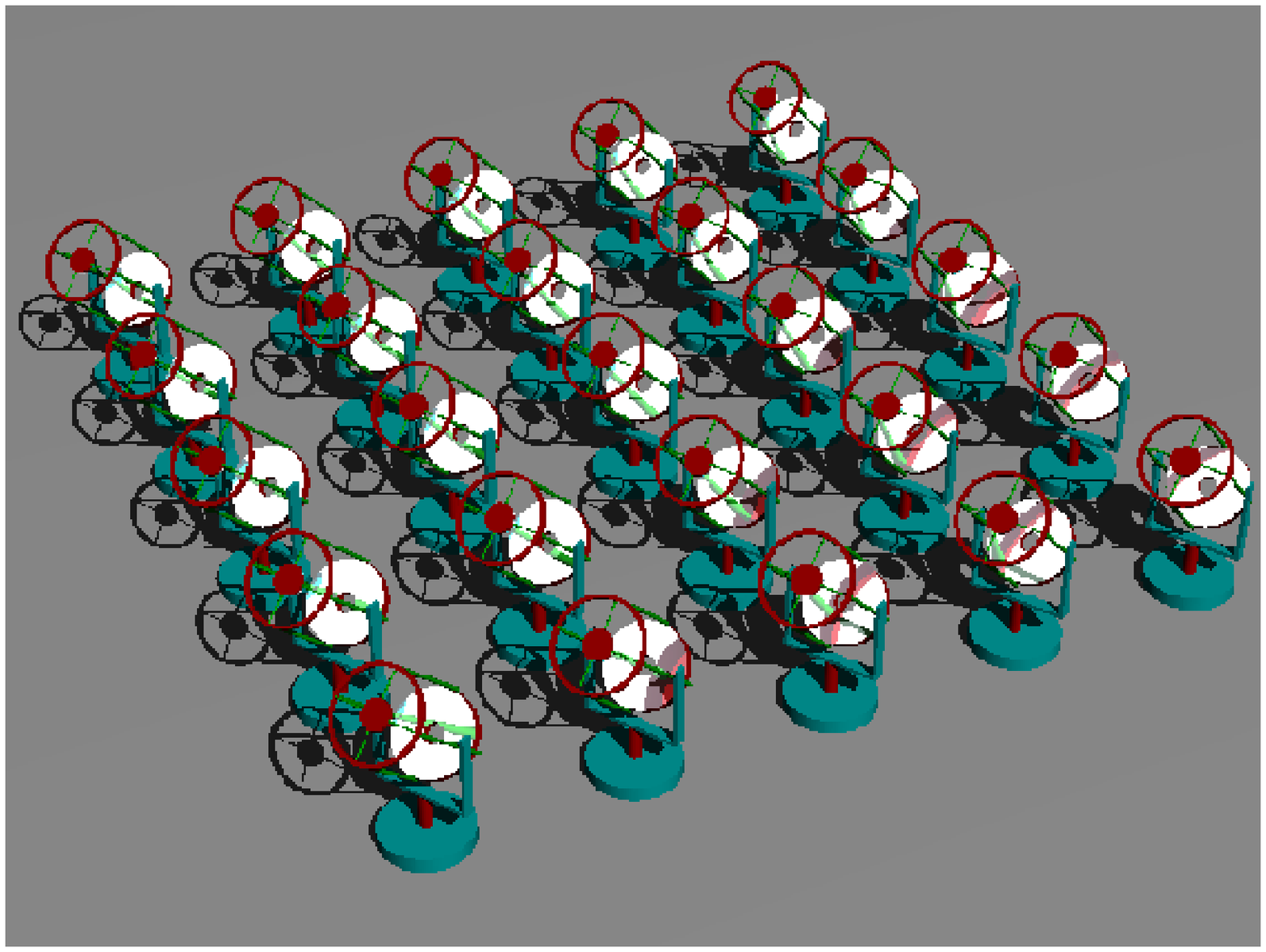,width=0.8 \linewidth}
\label{fig:array}
\end{figure}

\newpage
\tableofcontents
\newpage
\listoffigures

\newpage

\section{Introduction}

Imaging surveys are limited by depth, angular coverage and angular resolution.
There are currently several proposals for wide field telescopes
and instrumentation which promise great gains in the first two
of these variables 
(the CFHT MegaPrime Project (\citeNP{bvc+98}; \citeNP{cfhtmegaprimewebsite});
Megacam on the MMT (\citeNP{wrw+98}; \citeNP{mgg+98}; \citeNP{ga98}; \citeNP{mmtmegacamwebsite});
the UK `Vista' project \cite{vistawebsite}; 
the `Dark Matter Telescope' \cite{dmtelescopewebsite};
Suprime-Cam for Subaru \cite{suprimecamwebsite};
Omega-Cam for the ESO VST at Paranal \cite{omegacamwebsite};
the Canadian CFHT 8m upgrade proposal).  
Unfortunately, these designs are hampered by the limited angular resolution
available from the ground; 
at $m \sim 25$ most faint galaxies are poorly resolved at
even the best sites, and we know from e.g.~the Hubble Deep Field
that galaxies become still smaller as one pushes fainter, and there
is a wealth of data lying tantalizingly beyond the resolution
of conventional ground-based telescopes.  

Atmospheric seeing arises from spatial fluctuations in the 
refractive index associated with
turbulent mixing of air with inhomogeneous entropy and/or water vapor content 
(e.g~\citeNP{roddier81}).
High order adaptive optics (AO) can achieve spectacular 
improvement in angular resolution
on large telescopes (see e.g.~the reviews of \citeNP{beckers93}; \citeNP{roddier99}), 
but has not been applied to wide-field imaging due to the limited `isoplanatic 
angle' this being the angular distance around  the guide star within
which target objects sample effectively the same refractive index
fluctuations.
There have been discussions of `multi-conjugate' systems 
to increase the field of view (e.g.~\citeN{fl85}), 
but little concrete has yet to emerge from this.  
Here we shall explore the possibility of
of achieving a more modest but still valuable gain in resolution 
by using an array of small telescopes with fast guiding or `tip-tilt'
correction.  In what follows we will first review why one
would want to use tit-tilt on small telescopes, we then discuss the
`isoplanatic angle' problem for fast-guiding, and how this can
be overcome using multiple telescopes and new technology in the
form of `orthogonal transfer' CCD technology.

\begin{figure}[htbp!]
\centering\epsfig{file=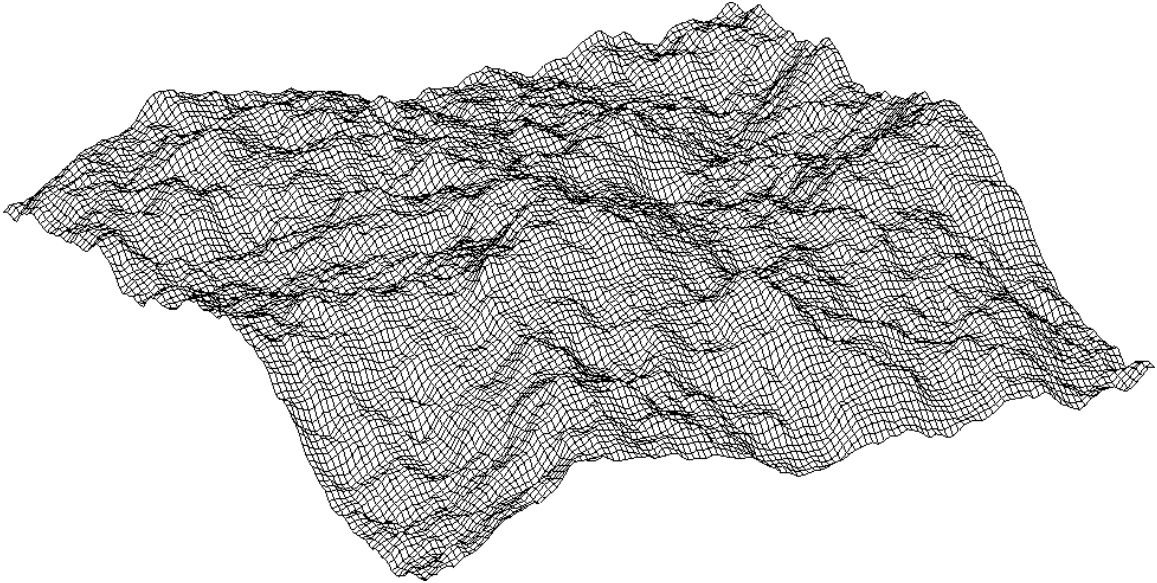,width=\figwidth}
\caption[Realization of wavefront deformation due to Kolmogorov turbulence.]
{A realization of the phase 
fluctuation $\varphi(\br)$ produced by atmospheric turbulence with 
a $P(k) \propto k^{-11/3}$ Kolmogorov law power spectrum. This is the
form of an initially flat wavefront from a distant source after
it has propagated thought the atmosphere.  No scales are shown for the
axes because the power-law power spectrum gives rise to a `scale-invariant'
statistical topography; a blow-up of a region of this realization is, aside
from a scaling of amplitude, statistically indistinguishable from the original,
}
\label{fig:phaseplot}
\end{figure}

Fast guiding is a common feature of modern large telescope
designs, and can be quite useful for dealing with `wind shake' or
other local sources of image wobble.  However, 
for realistic turbulence spectra, and for most sensible measures
of image quality, fast guiding has relatively
little effect on the {\it atmospheric\/} contribution 
to seeing for large telescopes.
For fully developed Kolmogorov turbulence 
(e.g.~\citeNP{tatarski61}) the structure function for
phase fluctuations is $S_\varphi(r) \equiv 
\langle (\varphi(r) - \varphi(0))^2 \rangle = 6.88 (r / r_0)^{5/3}$.
This says that the rms phase difference between two points grows in 
proportion to the $5/6$ power of their separation.  The character of the
phase fluctuations imposed on wavefronts is shown in figure \ref{fig:phaseplot}.
The amplitude of the phase fluctuations is characterized by the
`Fried length' $r_0$ \cite{fried65}, which is the separation for which
the rms phase difference is of order unity (actually $\sqrt{6.88}$ radians), and
is on the order of 20cm at good sites in the visible).

The rapidly growing amplitude of the structure function means
that the phase variations are dominated by the lowest order modes.
For example, if we ignore piston, the phase 
variance averaged over a circular aperture
of diameter $D$ is $\langle\varphi^2\rangle = 1.013 (D / r_0)^{5/3}$
but this drops to $\langle\varphi^2\rangle = 0.130 (D / r_0)^{5/3}$
if the lowest order Zernike modes of tip and tilt are removed \cite{fried65}.
Thus applying tip-tilt correction reduces the phase variance by
a factor $1.013 / 0.130 = 7.8$ which is both substantial and independent
of the diameter of the telescope.  We can safely conclude from this that
a telescope with $D$ less than a few times $r_0$ will, after tip-tilt
correction, have residual phase variance which is small compared to unity
and will therefore give close to diffraction limited performance.  

What about larger telescopes? If $D / r_0 \gg $ a few then the residual
phase variation is large compared to unity, so such telescopes will not
be diffraction limited.  As with a small telescope, the primary effect of tip-tilt
is to reduce the phase fluctuations on scales of order the telescope
diameter.  This will cause a dramatic improvement in the transmission of the
telescope for frequencies approaching the diffraction limit, but the uncorrected
transmission for such frequencies is essentially zero, so even a large
gain here does little good.  There is some reduction in phase variations
on smaller scales --- separations on the order of $r_0$ that is --- with some
corresponding increase in useful image quality which we can estimate as
follows:  The root mean squared tilt of the wavefront averaged over
the aperture is on the
order of $\delta \theta \sim \delta \varphi \lambda / D
\sim \sqrt{S_\varphi(D)} \lambda / D 
\sim \lambda / (D^{1/6} r_0^{5/6})$.  More precisely, 
we find the variation in position
the uncorrected PSF centroid to be a Gaussian $\exp(-\theta^2 / 2 \sigma_{\rm tilt}^2)$
with $\sigma_{\rm tilt}^2 = 0.169 (\lambda / r_0)^2 (r_0 / D)^{1/3}$
whereas the uncorrected PSF has FWHM $= 1.0 \lambda / r_0$ for
$D \gg r_0$, which is the same as for a Gaussian with variance
$\sigma_{\rm total}^2 = 0.181 (\lambda / r_0)^2$.  To a crude approximation,
which actually becomes quite good for $D \gg r_0$, one might expect the
corrected PSF to approximate a Gaussian with $\sigma^2 = \sigma_{\rm total}^2
- \sigma_{\rm tilt}^2$ with corresponding improvement in
image quality (which we take to be the inverse of the area of the PSF) of
\begineq
\label{eq:imagequalityvsD}
{\sigma_{\rm total}^2 \over \sigma_{\rm total}^2 - \sigma_{\rm tilt}^2}
\simeq {1 \over 1 - 0.93 (D / r_0)^{-1/3}}
\endeq
so the gain from tip/tilt is predicted to decrease, albeit somewhat
slowly, for large $D / r_0$.
This theoretical expectation \cite{fried66} has been 
widely discussed and studied in detail
(\citeNP{young74}; \citeNP{christou91}; 
\citeNP{glindemann97}; \citeNP{jenkins98})
and it turns out that, for a filled aperture, a pupil diameter $D \simeq 4 r_0$
maximizes the normalized Strehl ratio,
this being defined as the central value of the normalized PSF as compared to that
for a large telescope, and the gain for $D=4r_0$ is a factor $\simeq 4.0$.
For $D/r_0 = 10$ the gain is $\simeq 2.0$ and for $D/r_0 = 50$ the gain is
a factor $1.4$.  These latter numbers are in quite good agreement
with the crude estimate (\ref{eq:imagequalityvsD}).
This expectation has also been confirmed in practice by 
\citeN{mfa+91} who used HRCAM on the CFHT with the pupil stopped down
to $D = 1.2$m,
and by \citeN{roddier92} with the UH adaptive optics system working in
tip/tilt mode again with the CFHT stopped down to 1m aperture.
These conclusions are somewhat dependent on the assumption of 
fully developed Kolmogorov turbulence. Recently
\citeN{mtz+98} have reported deviations from the $P(k) \propto k^{-11/3}$
law at La Silla which they characterize, in the context of the von-Karman model, as
an `outer-scale' of $\sim 20$m, and 
a number of the measurements reviewed by \citeN{azb+97}, have also  given
fairly small values. A finite value for the outer scale will tend to
further reduce the effect of tip-tilt correlation on large telescopes.

Another way to look at this problem is in terms of `speckles'.  A
snapshot of the PSF for a large telescope consists of a set of speckles,
each of which is about the size of the diffraction limited PSF, and
there being on the order of $N \sim D^2 / r_0^2$ speckles in total, i.e.~on
the order of the number of $\sim r_0$ sized patches within the pupil.  These speckles
dance around on the focal plane 
(see \href{http://www.ifa.hawaii.edu/~kaiser/wfhri}{\tt http://www.ifa.hawaii.edu/$\sim$kaiser/wfhri}
for an animated movie showing the evolution of PSFs for a range of 
telescope diameters).  For $D / r_0 \simeq 4$
it is found that much of the time a substantial fraction of
the light (say 25\% or so) is in a single bright central speckle, and
by tracking the centroid --- or better still tracking the peak of the
brightest speckle \cite{christou91} --- one can keep this 
bright dominant speckle at a fixed point on the focal plane, resulting in a
PSF with a diffraction limited core component.

For Kolmogorov turbulence the seeing angle (defined as the FWHM of the PSF) is
$\Delta\theta = 0''.5 (\lambda / 0.5 \mu{\rm m}) (r_0(\lambda)/ 20{\rm cm})^{-1}$
with $r_0(\lambda) \propto \lambda^{6/5}$.
Analysis of SCIDAR measurements at Mauna Kea by \citeN{racine96}
gave a median $\Delta \theta = 0''.43$ at $\lambda = 0.5 \mu{\rm m}$
corresponding to $r_0(\lambda = 0.5 \mu{\rm m}) = 22.6$cm
or, in the I-band, $r_0(\lambda = 0.8 \mu{\rm m}) = 40.0$cm
for which the optimal telescope diameter is then $D = 1.6$m.
A similar value of the mean free atmosphere 
Fried length ($r_0 = 30$cm at $\lambda = 0.55\mu{\rm m}$) was inferred
by \citeN{cs88} from correlation of the modulation
transfer function for wide binary stars of various separations.
This would suggest a 20\% larger optimal diameter, but
the result is dependent on their assumed model for the vertical
distribution of seeing.
Racine also found an approximately log-normal distribution of seeing angle
with 1-sigma points of $\Delta \theta = 0''.29, \; 0''.66$
corresponding to a range of $r_0(\lambda = 0.5 \mu{\rm m}) = 33.9, \; 15.1$cm,
and to a range in wavelength for which $r_0$ is
optimally matched to a 1.6m telescope of $\lambda(r_0 = 40{\rm cm}) = 0.57, 1.12 \mu$m.
Thus in the normal range of seeing conditions a $1.6$m diameter
fast guiding telescope could operate optimally over the range of passbands V,R,I, Z.
Note that in a multi-color survey performed in this way the resolution for the different passbands
scales as FWHM $\propto \lambda$ rather than FWHM $\propto \lambda^{-1/5}$
as in a conventional survey.

Applying fast image-motion correction to a small telescope should therefore
greatly enhance image quality, but only over a limited distance from the
guide star used to measure the motion.
Studies of the atmosphere using SCIDAR and  thermosonde probes
(\citeNP{bcp+76}; \citeNP{bcp+77};
\citeNP{rcg+90}; \citeNP{rng+93}; \citeNP{epr+94}; \citeNP{racine96}; \citeNP{avc98})
have shown that the source of seeing 
(conventionally characterized by $C_n^2(h)$, 
the intensity of the power spectrum of
refractive index fluctuations \cite{roddier81}) is highly structured and
stratified.  In the typical situation there is a substantial contribution
from very low level turbulence --- the planetary boundary layer and
`dome seeing' ---
but there are also comparable contributions from higher altitude
layers at $h \sim 1-10$km. The indications are that the
layers are thin with thickness on the order of 100m or so.
The low level turbulence
causes a coherent motion of all objects in the field, and is
relatively easy to deal with.  The higher altitude seeing is
more problematic, since the angular scale over which images move coherently --- the
`isokinetic' angle --- is on the order of $D/h$.  For $h = 5$km say
and $D = 1.5$m this is roughly 1 arc-minute.  It is perhaps worth mentioning
at this point that the \citeN{mfa+91} experiment seemed to show a
substantially larger isokinetic scale than the simple estimate given here.  
This is encouraging, as it would indicate a predominance of low-level
turbulence,
which would make our job easier, but it may
well be that they were lucky and observed at a time when the
high altitude seeing contribution was relatively quiescent.

The limited isokinetic angle has serious implications for wide field imaging.
For a 1-degree field say, the focal plane will consist of $\sim 3000$
isokinetic patches  moving independently, so one needs some
kind of `rubber focal plane' detector to track these motions.
Moreover, at high galactic latitudes at least, the mean separation
of stars which are sufficiently bright to guide on ($m \sim 16$) is on the order of
$4'$, so there are two few guide stars with which to determine the
deflection field at all points on the focal plane.

One way to avoid the latter problem would be to work at lower galactic latitude,
where bright guide stars are more abundant,
or to peer through globular clusters, but these approaches seem rather unsatisfactory.
The solution to these problems that we propose here has two key features.
The first is to use OTCCD \cite{tbs97} technology to implement the 
rubber focal plane. The second is to use an array of telescopes to
provide multiple samples of the atmospheric turbulence to provide the
information needed to accurately guide out image motion at all points in the field of view.

In an OTCCD device, as in an ordinary CCD, the electrons created by impinging photons
are are trapped in a grid of potential wells.  The difference is that the origin
of the grid can be shifted with respect to the
physical pixels by fractional pixel displacements, and the accumulating
charge can therefore be moved around 
quasi-continuously, and in multiple directions, to accommodate drifting of the
images due to the atmosphere.  A camera made of a large number
of such devices could then shift charge on different parts of the focal
plane independently.

\begin{figure}[htbp!]
\centering\epsfig{file=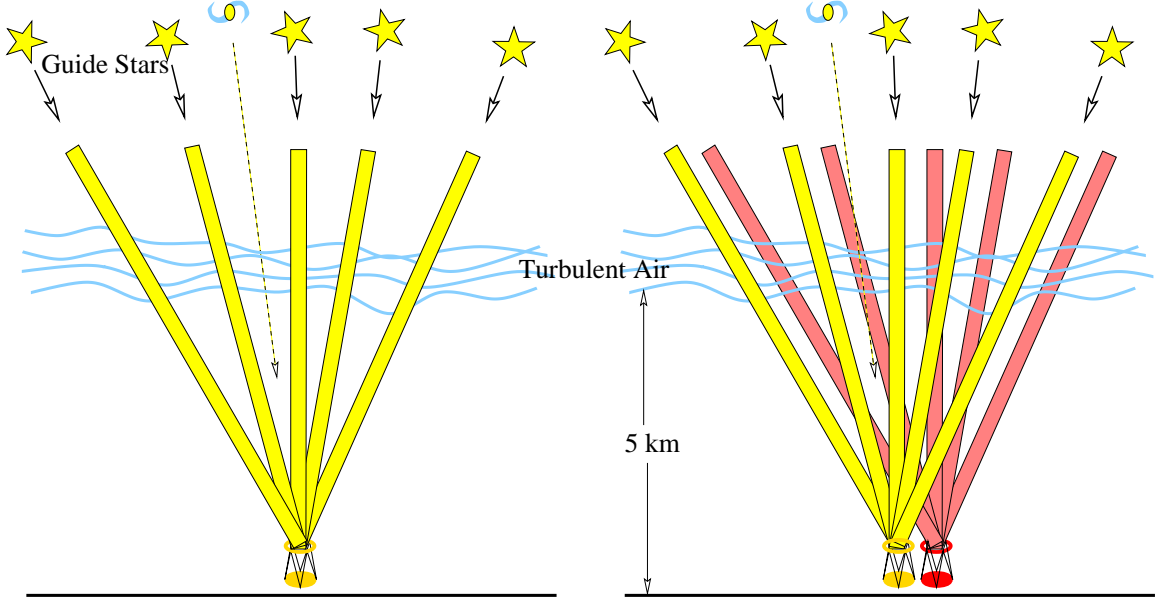,width=\figwidth}
\caption[Schematic illustration of multiple sampling of turbulent layer]
{This figure shows schematically how an array of telescopes can
be used to obtain effectively complete sampling of a high-altitude
turbulent layer.}
\label{fig:beamsplot}
\end{figure}

To see how an array of telescopes might solve the problem of limited
guide stars consider the simple case of
a single thin layer of high-altitude turbulence at height $h$
above the telescope.  
A single small telescope monitoring a set
of guide stars will provide a set of samples of the
the image deflection field, scattered over a region of size $\sim h \Theta$
where $\Theta$ is the angular field size,
but with spacing somewhat larger than the deflection coherence scale.  
A neighboring telescope observing the
same set of stars will provide another set of samples of the
deflection field with the same pattern as the first, but displaced by the
vector separation of the telescopes
as illustrated in figure \ref{fig:beamsplot}.  With an array of telescopes one can further
increase the density of sampling of the deflection field until one has
full coverage.  In this scheme then, the information needed to guide out the
motion of a target object image seen through some patch of the turbulent layer
by a particular telescope
would be provided by one or more other telescopes in the array which are viewing
bright guide stars through the same patch of turbulence.

A single thin layer of turbulence is something of an idealization, and with
multiple or thick layers one clearly needs more information.  However,
there {\sl is\/} much more information at our disposal: a generic feature of 
Kolmogorov turbulence is that the turnover time for small scale eddies is
long as compared to the time-scale for winds to convect the eddies
through their length, so to some approximation we can adopt the `frozen turbulence'
assumption and use the additional information from
positions of guide stars viewed at earlier times at points upwind of the
point in question to constrain the deflection field.

Consider then an array of perhaps a few tens of
small telescopes each acting as an incoherent detector --- there being
no attempt here to co-phase the signals from separate telescopes as is done with
a interferometer array --- but sharing the image motion data needed to
implement low order AO in the form of fast guiding.  Such an array,
monitoring of order several hundred guide stars (for a nominal 1-degree field say)
would provide many thousands of skewer like samples through the
layers of turbulence flowing over the array.  How though is one to make
sense of this huge torrent of data in practice?
We believe the answer is to exploit the statistically Gaussian nature of the
turbulent layers.  The eddy size which dominates the deflection here is on
the order of the telescope diameter $D$.  Assuming that the thickness of the turbulent
layer exceeds this then the central limit theorem effectively guarantees
that the phase perturbation $\varphi$ imposed on a wavefront passing through such a
layer should be a statistically homogeneous and isotropic (though
flowing) Gaussian random field. For fully developed Kolmogorov turbulence the power spectrum is
$P_\varphi(k) \propto k^{-11/3}$ though this may flatten at low wave-number
to something like the von Karman form $P(k) \propto (k^2 + k_0^2)^{-11/6}$
parameterized by $k_0 \equiv 2 \pi / \lambda_0$.
In reality, atmospheric turbulence is intermittent, with the strength of the
turbulence varying on time-scales of tens of minutes \cite{racine96}, and this will
break the very large-scale spatial homogeneity, but 
the process may well be effectively stationary on smaller length scales.

The deflection of the image centroid $\delta \vtheta$ is, as we shall see below, 
obtained by taking the derivative of the phase perturbation and averaging over the telescope
pupil.  This is a linear
function of the phase fluctuation field and so should also have
Gaussian statistics, so
this allows one to write down the joint probability
distribution for a set of $N$ deflections $\delta \theta_I$ where $I$ is
a compound index specifying the object, the telescope, the time of observation,
and also the Cartesian component of the deflection:
\begineq
\label{eq:multivariatepdf}
p(\delta \theta_0, \delta \theta_1, \ldots) d^N \delta \theta
= {1 \over \sqrt{(2 \pi)^N |M|}} \exp(-\delta \theta_I M_{IJ}^{-1} \delta \theta_J / 2)
\endeq
where the covariance matrix is 
\begineq
M_{IJ} \equiv \langle \delta \theta_I \delta \theta_J \rangle .
\endeq
This covariance matrix is a smooth and well defined function of
the vector separation of the telescopes $\Delta \bx$; the vector angular separation of the
objects $\Delta \btheta$; and the time difference $\Delta t$.
For sufficiently large fields one can obtain sufficiently dense sampling
in $\Delta \btheta$ and
by integrating over several minutes, it should be possible to accurately
determine the deflection covariance function 
$\xi_{ij} (\Delta \bx, \Delta \btheta, \Delta t) = \langle \delta\theta_i \delta\theta_j' \rangle$.
From this covariance matrix
one can then compute the conditional probability for the deflection of
a target object viewed with a given telescope at the current instant of time $t_0$ given the measured
deflections of a set of guide stars for $t < t_0$.
From this conditional probability $p(\delta\btheta_{\rm target} | \{ \delta\btheta \}_{\rm guide\; stars})$
one can extract both the mean conditional deflection $\overline {\delta\btheta}$ --- which
is the signal one uses to guide out the target object motion --- and also the covariance matrix for
the errors in the guide signal $\sigma_{ij} = \langle (\delta\theta_i - \overline{\delta\theta_i})
(\delta\theta_j - \overline{\delta\theta_j}) \rangle$ which, as we shall see, allows one to compute the
final PSF and thus monitor the performance of the system.

To summarize, the concept which emerges is of an array of telescopes,
each equipped with its own wide field detector divided into a large number of segments, 
each of which is either continuously
monitoring the position of a guide star or integrating.
The guide star data is fed to a
multi-variate Gaussian `probability engine', which feeds back to the telescopes the
necessary information for moving the accumulated charge on each integrating segment of the detector.
As we shall see, under good conditions, such 
an instrument should allow $\simeq \FWHM$ FWHM image quality over large
fields; while only a modest --- roughly a factor 3 --- increase in resolution over conventional
telescopes we believe that this is well worth having as much of the information of
interest in faint galaxy studies lies at spatial frequencies tantalizingly
close to, but beyond the resolution attainable with a large aperture single mirror
telescope.  
A nice feature of this design is that
it scalable to arbitrarily large collecting area with cost proportional to
area. 

In the rest of this paper we will discuss in more detail
the practicality of this approach.  
In \S\ref{sec:psf} we present calculations of the PSF 
and optical transfer function (OTF) for fast guiding.
We present a
number of objective measures of the image quality which are relevant for
different types of observation.  We discuss the constraint on pixel size
and telescope design imposed by
the requirement that the image quality not be degraded by detector resolution.
We also quantify how the PSF degrades with distance from the guide star.  
In \S\ref{sec:guidestars} we consider the constraints imposed
by the limited numbers of sufficiently bright guide stars.  
In \S\ref{sec:correlations} we discuss the spatial and temporal correlations of
image motions. We outline our guiding algorithm and what constraints are imposed on the telescope array geometry.
We find that there are currently insufficient data on the detailed structure and
statistics of atmospheric turbulence to 
definitively determine the performance and optimize the design for the
type of system we have in mind, but we describe the kinds of experiments
that should be done to resolve this.
In \S\ref{sec:detectors}
we discuss the OTCCD `rubber focal plane' detector.
We consider the costs of software development in \S\ref{sec:software},
and summarize the overall system cost in \S\ref{sec:costs}.
In \S\ref{sec:science} we outline some of the scientific opportunities that
this kind of instrument makes possible.

\section{Image Quality with Fast Guiding}
\label{sec:psf}

According to elementary diffraction theory (e.g.~\citeNP{bw64}),
the electric field amplitude at some position $\bx_{\rm phys}$ on 
the focal plane of a telescope
is the Fourier transform of the product of the telescope pupil
function $T(\br)$ with the atmospheric phase factor $e^{i\varphi(\br)}$, so
$E(\bx_{\rm phys}) \propto \int d^2 r\; T(\br) 
e^{i\varphi(\br)} e^{2 \pi i \bx_{\rm phys} \cdot \br / L \lambda}$.
Here $L$ is the focal length and $\lambda$ is the wavelength.
Squaring this gives the intensity which, suitably normalized, is the 
PSF $g(\bx_{\rm phys})$.
In what follows it is convenient to work in rescaled focal plane coordinates
$\bx \equiv 2 \pi \bx_{\rm phys} / L \lambda$.
The PSF is then the inverse Fourier transform of the OTF
\begin{equation}
\label{eq:psffromotf}
g(\bx) = \int {d^2z \over (2 \pi)^2} \; e^{-i\bx \cdot \bz} \tilde g(\bz)
\end{equation}
where
\begin{equation}
\label{eq:otf0}
\tilde g(\bz) = \int d^2r \; T(\br) T(\br + \bz) e^{i[\varphi(\br) - \varphi(\br + \bz)]}
\end{equation}
where we have normalized the pupil function so that $\int d^2 r \; T^2(\br) = 1$.
These results are valid in the `near-field' limit, which should be quite accurate for our purposes
\cite{rr86}.

In an idealized fast guiding telescope, the instantaneous PSF is measured from 
a bright `guide star', and its position is determined and used to guide the telescope.
In this section our goal is to determine the final corrected PSF averaged over
a long integration time.
Since the statistical properties of the phase fluctuation field $\varphi(\br)$
are given by Kolmogorov theory this is a well posed problem.  It is however somewhat complicated,
and the details of the PSF depend on the method used to determine the
center.  
We first review the calculation of the `natural' or uncorrected PSF in \S\ref{subsec:naturalpsf}. 
We discuss the approximation to the corrected OTF given by \cite{fried66} in \S\ref{subsec:friedmodel}.  
In \S\ref{subsec:centroidpsf} we compute the OTF and PSF for the case of guiding on the image centroid.
In \S\ref{subsec:isoplanatism} we show how the PSF for centroid guiding depends on the distance from the
guide star.
In \S\ref{subsec:alternativecentering} we discuss alternatives to the image centroid such as 
peak tracking, which yield somewhat superior image quality. 
Finally, in \S\ref{subsec:pixelization} we consider the effect of finite pixel size.

\subsection{The Natural PSF}
\label{subsec:naturalpsf}

The long-exposure uncorrected OTF was first given in the classic paper of
\citeN{fried66} and is obtained by taking the 
time average of (\ref{eq:otf0}). This requires the average of the complex exponential of the
phase difference $\psi(\br,\bz)  = \varphi(\br) - \varphi(\br + \bz)$.  
At fixed $\br,\bz$ this is a stationary (in time) Gaussian random process with 
with probability distribution 
$p(\psi) = (2 \pi \langle \psi^2 \rangle)^{-1/2} \exp(- \psi^2 / 2 \langle \psi^2 \rangle)$
and so the time average of the complex exponential is
\begin{equation}
\label{eq:expavg}
\langle e^{i \psi} \rangle = \int d\psi \; p(\psi) e^{i \psi} =
\exp(-\langle \psi^2 \rangle / 2).
\end{equation}
where the final equality follows on integrating by completing the square.
Now since the phase fluctuation field is also a statistically spatially homogeneous 
process, the phase difference variance or `structure function'
$\langle \psi^2 \rangle = \langle (\varphi(\br) - \varphi(\br + \bz))^2 \rangle$
is a function only of the separation of the
points: $\langle \psi(\br, \bz)^2 \rangle = S(z)$, and so in this special case the
OTF factorizes into the product of the diffraction limited
OTF $\tilde g_{\rm diff}(\bz) = \int d^2r \; T(\br) T(\br + \bz)$ and the
`atmospheric transfer function' $\tilde g_{\rm atmo}(z) = \exp(- S(z)/2)$.
For fully developed Kolmogorov turbulence the structure function is a power law
$S(z) \propto z^{5/3}$ and the OTF therefore has the form 
$\tilde g(z) \propto \exp(-\alpha z^{5/3})$.

\subsection{The Fried Model}
\label{subsec:friedmodel}

\begin{figure}[htbp!]
\centering\epsfig{file=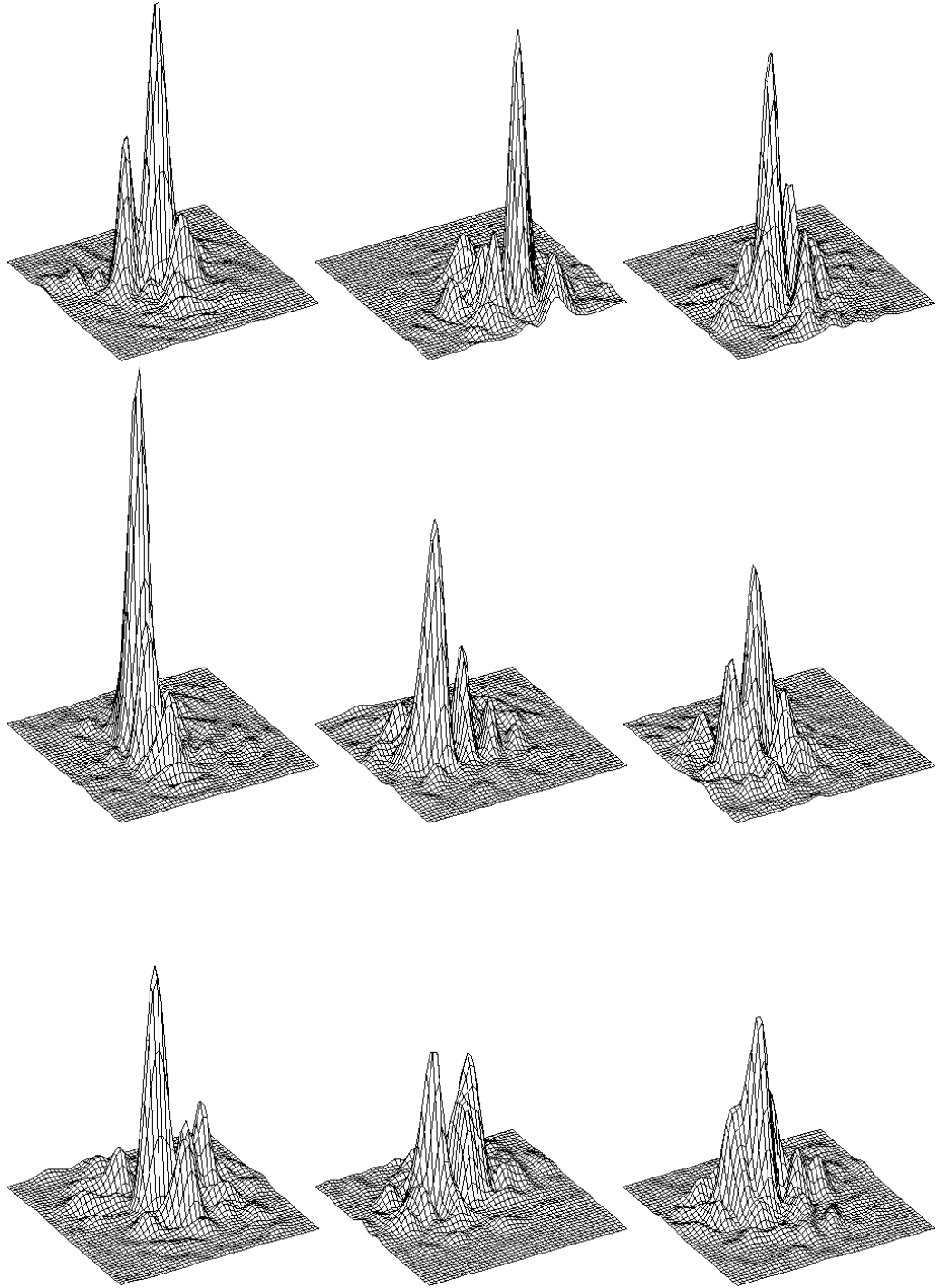,width=0.8\linewidth}
\caption[Numerical realizations of PSFs]
{Numerical realizations of some instantaneous PSFs for a 
telescope with $D/r_0 = 4$
generated as described in \S\ref{subsec:alternativecentering}.
The box size here is $1''.2 (\lambda / 0.8\mu{\rm m})$,
and the surfaces are 
normalized such that $\int d^2x\; g(x) = 1$.
These show graphically how the small telescope PSF typically contains a
very sharp spike, which is effectively diffraction limited, but that this
spike undergoes  large random displacements.  
}
\label{fig:psfexamples}
\end{figure}

A large part of the width of the natural PSF can be attributed to `wandering' of the
instantaneous PSF.  This is illustrated in figure \ref{fig:psfexamples}
which shows a set of realizations of the instantaneous PSF for a telescope
with $D/r_0 = 4$.  A series of animated images showing the continuous
evolution of atmosphere limited PSFs can be viewed at
\href{http://www.ifa.hawaii.edu/~kaiser/wfhri}{\tt http://www.ifa.hawaii.edu/$\sim$kaiser/wfhri}.
Fried argued that the corrected or `short-exposure' OTF,
i.e.~that obtained after taking out any net shift in these
instantaneous images before temporal averaging, should be of the form
\begineq
\label{eq:friedotf}
\tilde g_c(z) \propto \exp(-\alpha z^{5/3} + \beta z^2).
\endeq  
This result has a very simple and intuitively reasonable physical
interpretation: think of the uncorrected PSF as the convolution of
the short exposure PSF with the distribution $p(\overline{\bx})$ of the image shifts 
caused by any net tilt to the incoming wavefronts.
For steady turbulence this distribution is a Gaussian, 
so in Fourier space its transform is also a Gaussian, and so the 
short exposure OTF should be the long-exposure form divided by
a Gaussian which, for suitably chosen $\alpha$, $\beta$, is exactly what equation (\ref{eq:friedotf})
states.  The flaw in this argument --- as acknowledged by Fried in
a footnote --- is that it assumes that the image shift is statistically independent of
the other components of the wavefront distortion, which is not strictly
correct (though for some purposes it is a pretty good model).
Another limitation of Fried's analysis is that it identifies the
image shift with the tip and tilt Zernike coefficients of the
wavefront.  While this is qualitatively correct, the shift of the
centroid of the image --- which is the quantity most readily
measured in the type of system considered here --- differs somewhat from
the tip/tilt coefficients.
This problem has been reconsidered by several authors 
(\citeNP{young74}; \citeNP{christou91}; 
\citeNP{glindemann97}; \citeNP{jenkins98}) 
using a variety of approximations and/or simulation techniques. 
We now present a simple analytic calculation of the OTF and PSF for
fast centroid guiding.

\subsection{PSF for Centroid Guiding}
\label{subsec:centroidpsf}

The photon weighted centroid is defined as
\begineq
\label{eq:centroiddefinition}
\overline \bx \equiv \int d^2 x \;  \bx g(\bx).
\endeq
Now since $\tilde g(\bz) = \int d^2 x \; g(\bx) e^{i\bk\cdot\bz}$, 
the gradient of the instantaneous OTF is 
$\nabla \tilde g(\bz) = 
i \int d^2 x\; \bx g(x) e^{i\bx \cdot \bz}$,
so the instantaneous centroid is given by the gradient of the OTF at the origin:
\begin{equation}
\label{eq:centroid}
\overline \bx
= - i \nabla \tilde g(0) = - \int d^2r \; T^2(\br) \nabla \varphi(\br) 
= \int d^2r \; \bW(\br) \varphi(\br)
\end{equation}
where the second equality is obtained by direct differentiation of (\ref{eq:otf0}),
and the final result follows on integrating by parts and defining
the vector valued function $\bW(\br) \equiv \nabla T^2(\br)$.  The centroid is the
average of the wavefront slope weighted by $T^2$ \cite{glindemann97}; the
so-called `G-tilt'.
Note that this is not quite the same as the `Z-tilt' defined as the
tip/tilt components of the Zernike decomposition of the wavefront;
for a simple filled disk pupil function
the Z-tilt coefficients are the integral of $T^2(\br) \br$ times $\varphi(\br)$
whereas in
(\ref{eq:centroid}) the function $\bW(\br)$ is non-zero only at the pupil edge.
For low spatial frequency phase fluctuations the wavefront tip-tilt
coefficients and the centroid are effectively identical, but they
couple to high spatial frequency fluctuations rather differently.
A key feature of the centroid is that it is a {\sl linear\/} function of the random
phase fluctuation field, a fact which greatly facilitates the following calculation.

We shall need the covariance matrix for the centroid deflections, which,
from (\ref{eq:centroid}), is
\begineq
\label{eq:centroidcovariance0}
\sigma_{ij} \equiv \langle {\overline x}_i {\overline x}_j \rangle
= \int d^2r \int d^2 r' \; W_i(\br) W_j(\br) \langle \varphi(\br) \varphi(\br') \rangle
\endeq
and the trace of which gives the variance of the centroid.
For atmospheric turbulence
$\varphi(\br)$ is a statistically isotropic and homogeneous random field so
$\langle \varphi(\br') \varphi(\br)\rangle = \xi(\br' - \br)$ where
$\xi(\br)$ is the two-point function of the phase, and depends only
on $|\br|$. 
For Kolmogorov turbulence the phase two-point function is formally ill-defined (in
reality its value depends on the outer-scale cut-off) and it is more 
convenient to work with the
phase structure function 
$S(r) \equiv \langle (\varphi(\br') - \varphi(\br' + \br))^2 \rangle 
= 2 (\xi(0) - \xi(r))$ which 
is well defined and has a power-law form $S(r) = 6.88 (r / r_0)^{5/3}$, where $r_0$ is the
Fried length.  To evaluate (\ref{eq:centroidcovariance0}) then we replace
$\langle \varphi(\br) \varphi(\br') \rangle$ by $\xi(\br - \br') = \xi(0) - S(\br - \br') / 2$.
The dependence on the cut-off dependent but constant term $\xi(0)$ drops out, 
and in terms of the structure function $S(r)$ the centroid covariance matrix is then
\begineq
\label{eq:centroidcovariance}
\sigma_{ij} = - \half \int d^2r \; W_i(\br) (W_j \otimes S)_{\br}
\endeq
where we have defined the convolution operator $\otimes$ such that
$a \otimes b \equiv \int d^2 r' \; a(\br') b(\br - \br')$.
The centroid covariance matrix has dimensions of $1/L^2$.  For Kolmogorov
turbulence
\begineq
\sigma_{ij} = \alpha_{ij} D^{-1/3} r_0^{-5/3}
\endeq
where $\alpha_{ij}$ is a dimensionless matrix depending only on the
shape of the telescope input pupil.  For a circularly symmetric pupil
this matrix is diagonal and for a filled circular
aperture we find, from numerical integration, that
\begineq
\sigma_{ij} = 6.68 \delta_{ij} (D / r_0)^{-1/3} r_0^{-2}.
\endeq

The instantaneous centroid corrected PSF is, from (\ref{eq:otf0}),
\begin{equation}
g_{\rm c}(\bx) = g(\bx + \overline{\bx}) =
\int {d^2z \over (2 \pi)^2} \; e^{-i\bx \cdot \bz} 
e^{-i \bz \cdot \int d^2 r' \; \bW(\br') \varphi(\br')}
\int d^2r \; T(\br) T(\br + \bz) e^{i[\varphi(\br) - \varphi(\br + \bz)]}
\end{equation}
so the average centroid corrected OTF for a long exposure, which we shall denote by 
$\tilde g_c(\bz)$, can be written as
\begin{equation}
\label{eq:otf}
\tilde g_{\rm c}(\bz) =
\int d^2r \; T(\br) T(\br + \bz) \langle e^{i \psi(\br, \bz) } \rangle
\end{equation}
where 
\begin{equation}
\psi(\br, \bz) \equiv \varphi(\br) - \varphi(\br + \bz) - 
\bz \cdot \int d^2 r' \; \bW(\br') \varphi(\br')
\end{equation}
Just as before, for given $\br$, $\bz$, the phase factor $\psi(\br, \bz)$ is a stationary 
(in time) Gaussian random process so again
$\langle e^{i \psi} \rangle = \exp(-\langle \psi^2 \rangle / 2)$
but where now
\begin{equation}
\begin{matrix}{
\langle \psi(\br, \bz)^2 \rangle = 
S(\bz) - 2 \bz \cdot \int d^2 r' \; \bW(r') 
[\langle \varphi(\br') \varphi(\br)\rangle - \langle \varphi(\br') \varphi(\br + \bz)\rangle]
+ \bz \cdot \bsigma \cdot \bz \cr
= S(\bz) + \bz \cdot [(\bW \otimes S)_\br - (\bW \otimes S)_{\br + \bz}] + 
\bz \cdot \bsigma \cdot \bz
}\end{matrix}
\end{equation}
and so the corrected OTF is
\begin{equation}
\label{eq:fastguidingotf}
\tilde g_c(\bz) = \int d^2r \; T(\br) T(\br + \bz)
e^{ - [ S(\bz) + \bz \cdot [(\bW \otimes S)_\br - (\bW \otimes S)_{\br + \bz}] + 
\bz \cdot \bsigma \cdot \bz ] / 2}.
\end{equation}

This equation, along with (\ref{eq:centroidcovariance})and  the
definition $\bW(\br) \equiv \nabla T^2$, allows one
to compute the OTF $\tilde g_{\rm c}(\bz)$ for a given spectrum of phase fluctuations
$S(r)$ and pupil function $T(\br)$.
For isotropic turbulence, and for a circularly symmetric input pupil,
$g_c(\bz)$  is only a function of $|\bz|$, so one can take
$\bz$ to lie along the $x$-axis say in (\ref{eq:fastguidingotf}).
Note that the terms involving $\bW \otimes S$ in (\ref{eq:fastguidingotf}) are not independent
of $\br$ and so one cannot factorize the fast guiding OTF into a product of
$\tilde g_{\rm diff}$ and a purely atmospheric dependent term as was the case for
the uncorrected OTF.

\begin{figure}[htbp!]
\centering\epsfig{file=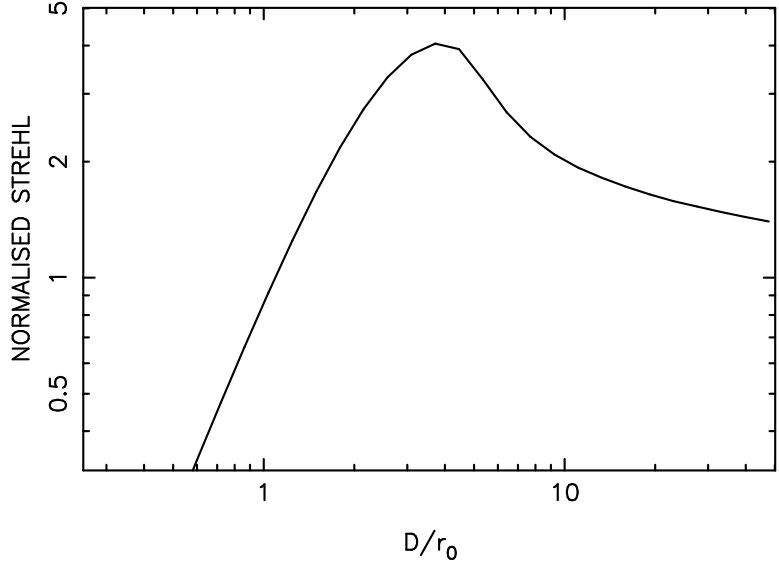,width=0.6 \linewidth}
\caption[Normalized Strehl ratio vs telescope diameter]
{The normalized Strehl ratio for fast centroid guiding is plotted  as
a function of $D/r_0$, the telescope diameter in units of the Fried length.
}
\label{fig:strehlplot}
\end{figure}

The `normalized Strehl ratio' is shown in figure \ref{fig:strehlplot}.  This is
the ratio of the central intensity of the normalized corrected PSF 
$g(\bx = 0) = (2 \pi)^{-2} \int d^2 z\; \tilde g(\bz)$
to that
for a very large telescope, and is a useful measure of the image quality.
This figure displays the well known result that according to this criterion
the best image quality is
obtained for telescope diameter $D \simeq 4 r_0$;
for smaller telescopes the seeing is limited by the size of the Airy disk
while for larger telescopes tip/tilt or fast guiding becomes ineffective at
reducing the phase variance.  

\begin{figure}[htbp!]
\centering\epsfig{file=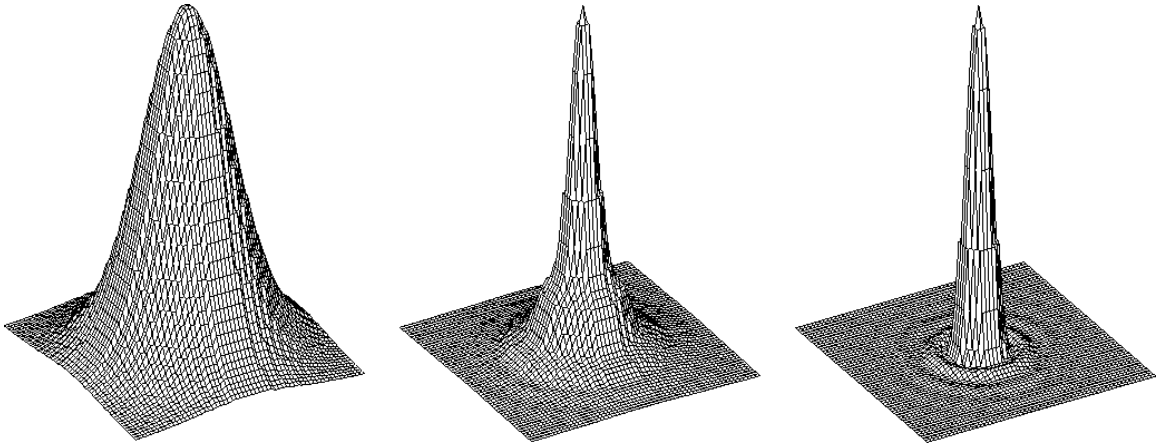,width=\figwidth}
\caption[Natural, fast guiding and diffraction limited PSFs.]
{Left hand panel shows the PSF for natural seeing. The
center panel shows the result of fast-guiding on a small
telescope.  The right hand panel shows the diffraction
limited PSF.
A wavelength $\lambda = 0.8\mu$m, a Fried length of
$40$cm and a telescope diameter of $D = 1.6$m were assumed.
The box side here is $1''.0$.  The plots
have been normalized to the same central value.}
\label{fig:psf3d}
\end{figure}

\begin{figure}[htbp!]
\centering\epsfig{file=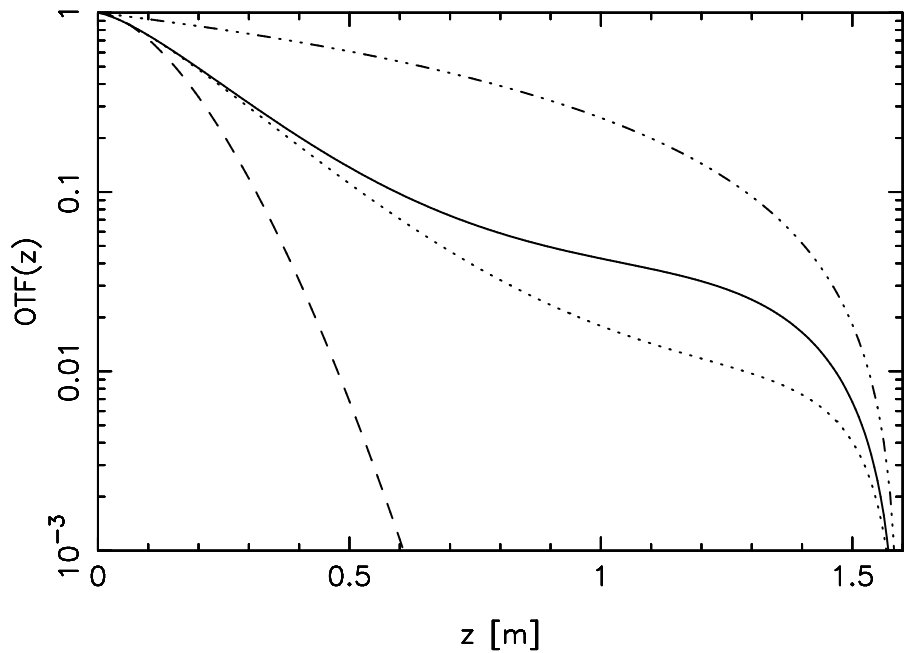,width={0.7 \linewidth}}
\caption[OTF for fast guiding]
{Optical transfer function for uncompensated
(dashed) and centroid guided (solid) images.  The dotted line is the Fried
approximation. 
The dot-dot-dot-dash line is the
OTF for a diffraction limited telescope. 
}
\label{fig:otf}
\end{figure}

\begin{figure}[htbp!]
\centering\epsfig{file=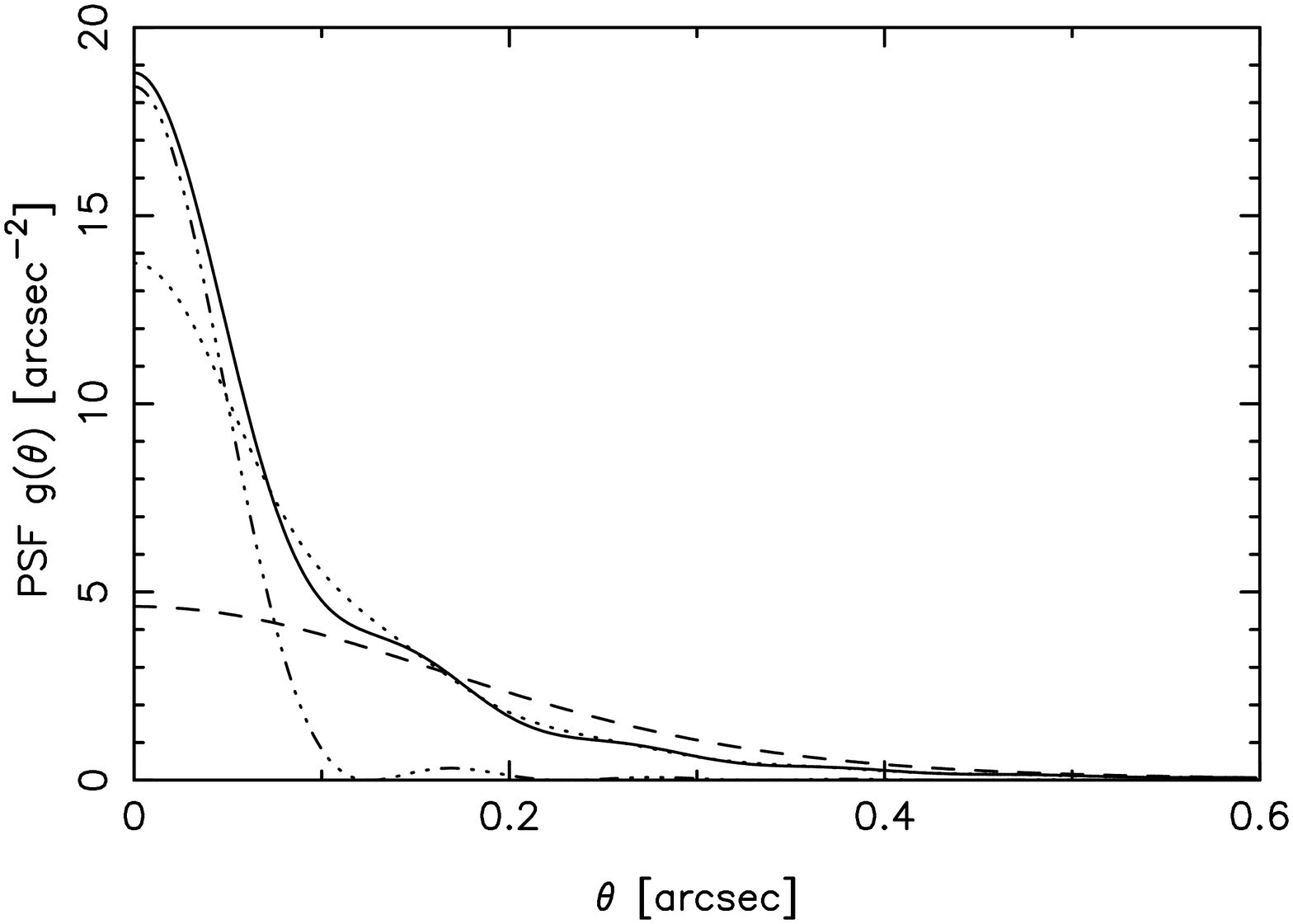,width={0.7 \linewidth}}
\caption[Radial profile of PSF for fast guiding]
{PSF obtained by Fourier
transforming the OTFs of figure \ref{fig:otf}.  
A wavelength $\lambda = 0.8\mu$m, a Fried length of
$40$cm and a telescope diameter of $D = 1.6$m were used.  
The curves are normalized so that $\int d^2 \theta \; g(\btheta) = 1$,
except for the diffraction limited case which has been multiplied
by $0.25$.
}
\label{fig:psf}
\end{figure}

The point spread function $g(\bx)$ computed as the Fourier transform of the OTF
given by equation (\ref{eq:fastguidingotf}) is shown in figure \ref{fig:psf3d} for fast guiding with
a telescope of optimal diameter $D = 4 r_0$.
Figure \ref{fig:otf} shows the OTF and figure \ref{fig:psf} shows the radial
profile of the PSF. To set the 
physical scales in these examples the Fried length was taken to
be $40$cm as appropriate for a good site like Mauna Kea at $\lambda \simeq 0.8 \mu$m
and which gives uncorrected FWHM $\simeq 0''.4$.  

There is no unique way to characterize the image quality of
a telescope.  It is clear from figure \ref{fig:otf} that the
gain in signal (and therefore in signal to noise) is
huge for frequencies approaching the diffraction limit of the
telescope, where the natural seeing OTF is exponentially suppressed.
Comparing the natural seeing and fast guiding PSFs we find:
\begin{itemize}
\item The normalized Strehl ratio is increased by a factor $4$.
\item The FWHM is reduced by a factor $3.1$ from $0''.4$ to $\FWHM$.
\item The resolution, according to the Rayleigh-style measure
	of the separation of a pair of equally bright stars which just
	produce separate maxima after convolution, is
	improved by a factor $3.44$ from $0''.324$ to $0''.094$.
\item The efficiency for detection of isolated point sources 
	against a sky background,
	which is proportional to $\langle g \rangle \equiv \int d^2 x\; g^2(\bx)$,
	is increased by a factor $2.16$.
\item The variance in position for a point source of flux $F$,
	detected as a peak of the image smoothed with the PSF, and seen
	on a noisy background with sky variance (per unit area)
	$\sigma^2$ is
	$$ \langle \Delta x^2 \rangle = 
	{ 8 \sigma^2 \int d^2 x | \nabla g |^2
	\over F^2 (\nabla^2 g(x = 0))^2} = { 8 \sigma^2 \over F^2}
	{\int {d^2 z \over (2 \pi)^2} z^2 \tilde g^2(z) \over
	\left[ {d^2 z \over (2 \pi)^2} z^2 \tilde g(z) \right]^2} $$
	and is decreased by a factor $\sim 130$.
\item	The efficiency for weak-lensing measurements is also
	increased by up to about a factor 120 for small galaxies
	as we show in more detail in \S\ref{subsec:lensing}.
\hide{
\item	The variance in the centroid of a bright object due to
	photon counting statistics (assuming the sky noise to
	be negligible) is proportional to 
	$\langle x^2 \rangle = \int d^2 x \; x^2 g(x)$ and is
	decreased by a factor  $1.45$. 
} 
\end{itemize}

It is apparent that there is spectacular improvement in the quality
of the core of the PSF. 
Crudely speaking, one can characterize the PSF as a near
diffraction limited core, which, for $D/ r_0 = 4$, contains about $1/4$ of the light,
superposed on an extended halo with width similar to the
uncorrected PSF. 
The low frequency `halo' can be removed
by spatial filtering, and images
with effectively diffraction limited resolution can thereby be
generated.

\subsection{Isoplanatism}
\label{subsec:isoplanatism}

\begin{figure}[htbp!]
\centering\epsfig{file=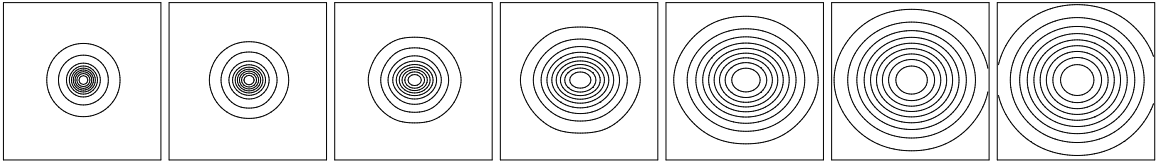,width=\figwidth}
\caption[Contour plots of PSFs as a function of guide star distance.]
{Contour plots of PSFs as a function of guide star distance.
These were calculated for a single layer of turbulence, a telescope
diameter $D = 1.6$m, and are for
target objects which sample the turbulent layer
at distances 0, 0.25m, 0.5m, 1m, 2m, 4m, 8m from the guide star.
These plots show the theoretically expected radial PSF anisotropy at
intermediate separations.
}
\label{fig:isoplanatism1}
\end{figure}

\begin{figure}[htbp!]
\centering\epsfig{file=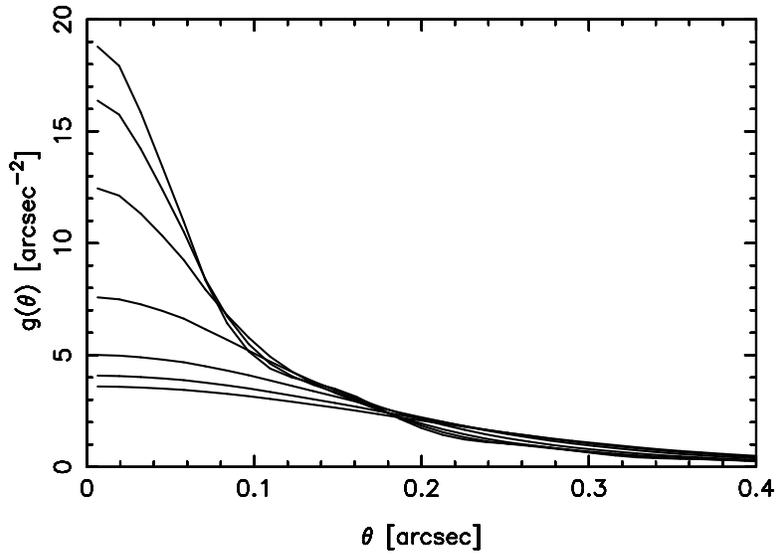,width= 0.6\linewidth}
\caption[Radially averaged profiles of PSFs as a function of guide star distance.]
{Radially averaged profiles of PSFs as a function of guide star distance.
These were calculated for a single layer of turbulence and are for
target objects which sample the turbulent layer
at distances 0, 0.25m, 0.5m, 1m, 2m, 4m, 8m from the guide star, just as in
figure \ref{fig:isoplanatism1}.
}
\label{fig:isoplanatism2}
\end{figure}

Equation (\ref{eq:fastguidingotf})
applies exactly only in the immediate vicinity of a guide star.  What happens if we
guide on the centroid of a certain star, but 
observe an object some finite angular distance 
$\Delta \btheta$ away?  
For a single deflecting screen at altitude $h$, 
equation (\ref{eq:fastguidingotf}) still holds,
but with the understanding that $\bf W(\br)$ be displaced 
from the origin by $\Delta \br = h \Delta \btheta$. 
(It is also relatively straightforward to generalize the
analysis here to allow for finite guide star sampling frequency, or
to guide using some linear combination of centroids of
a number of guide stars, but we shall not elaborate on that at this point).  
The result, for a range of distances from the guide star, is shown in
figures \ref{fig:isoplanatism1}, \ref{fig:isoplanatism2}.
It is interesting to note that if the range of the phase deflection correlations
is limited to some correlation scale length
$r_c$, as in the von Karman model for instance, so the structure
function becomes flat at $r \gg r_c$, then the terms involving $\bW \otimes S$
become negligible if we guide on a star which lies at distance far from the
target object and we find the simple and intuitively reasonable result that the
OTF is the product of the uncorrected OTF with 
$\exp(-\bz \cdot \bsigma \cdot \bz / 2)$ or
equivalently that the PSF is the convolution of the uncorrected PSF with
$\exp(- \bx \cdot \bsigma^{-1} \cdot \bx / 2)$ which is just the distribution of the
centroid deflections $p(\bx)$.

It is interesting to compare figure \ref{fig:isoplanatism1} with the
results of \cite{mfa+91}.  They measured shapes of several stars up to
$100''$ from the guide star and saw an increase in ellipticity
with distance but very little increase in size.  This suggests that their
$100''$ separation corresponds to a physical separation of $\simeq 0.25-0.5$m
and this would be consistent with a layer of turbulence at $h \simeq 1$km.

We have also compared the exact results obtained using (\ref{eq:fastguidingotf})
with the `Friedian' approximation (that the uncorrected PSF is the convolution of the
corrected PSF with $p(\bx)$), which is
\begineq
\label{eq:friedotf2}
\tilde g_{\rm Fried}(\bz) = 
e^{- \half (S(z) - \bz \cdot \bsigma \cdot \bz) } g_{\rm diff}(\bz).
\endeq
We find that the approximate  OTF agrees asymptotically with the exact
calculation for small argument, but there are sizeable departures at
large $z$, and consequently the high spatial frequency features of the PSF are
not accurately reproduced. 

\subsection{Alternative Guiding Schemes}
\label{subsec:alternativecentering}

In the previous sections we considered in detail the case of guiding on
the {\sl centroid\/}. This was largely for the sake of mathematical
convenience, as it allowed us to derive fairly simple but exact (at least
in the near field limit) formulae for the PSF and OTF, but these may not be optimal.
Alternatives to centroid guiding have been considered by \citeN{christou91}, who
has performed simulations to compare tip-tilt, centroid guiding and
also the so called `shift and add' or `peak tracking'
procedure where the image is centered on the peak of the brightest speckle.
By construction, peak tracking will optimize the Strehl ratio.
Other possibilities that will tend to give more weight to the central
parts of the PSF, and which might therefore be expected to
sharpen up the image quality near the peak, are to take the average of $x_i$, but weighted by
some non-linear function of the PSF.  For instance, one possibility would be
to define the image center as
\begineq
\overline \bx = \int d^2 x\; \bx g^2(\bx) / \int d^2 x\; g^2(\bx).
\endeq

\begin{figure}[htbp!]
\centering\epsfig{file=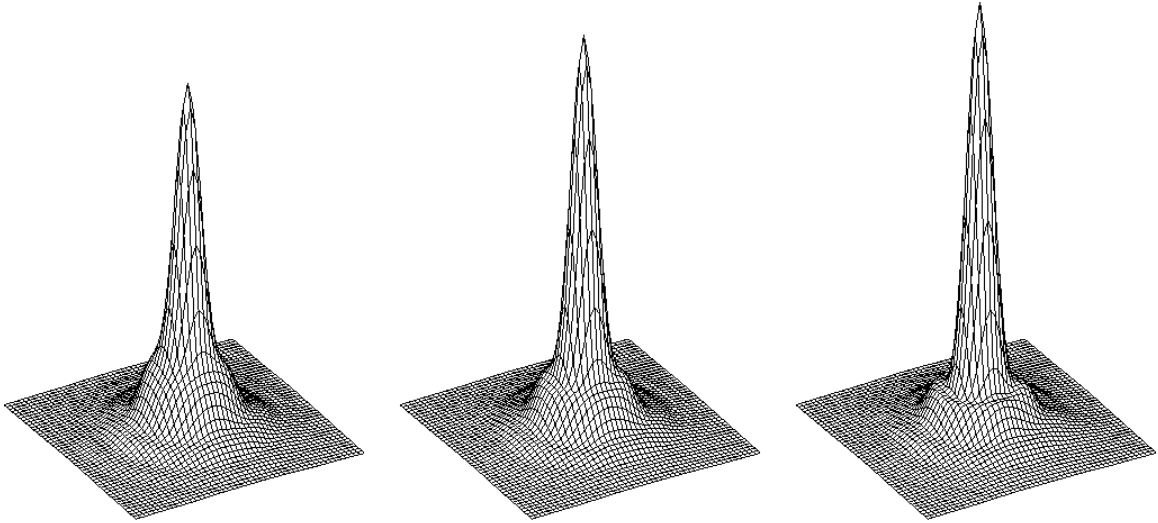,width=\figwidth}
\caption[3D surface plots of PSFs for alternative centering schemes]
{Surface plots of averaged PSFs computed numerically and re-centered
using a variety of algorithms.  On the left is the centroid. 
The center plot shows the result of centering on the $g^2$ weighted
centroid as described in the text. On the right is shown the
result for peak tracking.  The PSFs are normalized to equal volume
below the surface. The box size is $1''.0$.
}
\label{fig:altpsf_3D}
\end{figure}

\begin{figure}[htbp!]
\centering\epsfig{file=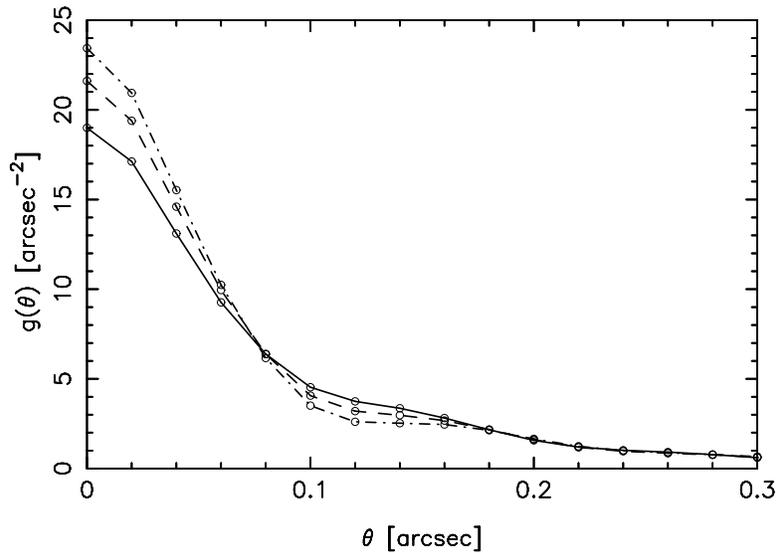,width=0.6 \linewidth}
\caption[Profiles of PSFs for alternative centering schemes]
{Radial profiles of simulated PSFs as shown in
figure \ref{fig:altpsf_3D}.  These show that the $g^2$ weighted centroid and
peak tracking give further improvements in the Strehl ratio 
of $\sim 15\%$, $25\%$ respectively as compared to
guiding on the centroid.
}
\label{fig:altpsf_pro}
\end{figure}

To explore the performance of these alternative centering algorithms --- which
are more difficult to treat analytically --- we have made simulations 
similar to those of Christou in which
we generate a large number of 
realizations of Gaussian random field phase screens with Kolmogorov spectrum
and compute the
integrals (\ref{eq:psffromotf}), (\ref{eq:otf0}) numerically 
to obtain realizations of the speckly
PSFs which we then re-center using various different algorithms and
then sum the result. Some example PSFs were shown in 
figure \ref{fig:psfexamples}. The result of averaging thousands
of such PSFs with various re-centering schemes are shown in figures 
\ref{fig:altpsf_3D} and \ref{fig:altpsf_pro}.
The result of this analysis is that with the more sophisticated
centering schemes considered here one can obtain a 15-25\%
improvement over centroid guiding and therefore an overall improvement in
normalized  Strehl ratio of about 5.

So far we have ignored the effect of read-noise and photon counting
uncertainty in the guide star position determination.  Of the
schemes considered here these are most problematic for the centroid,
and so noise considerations further favor peak tracking or some non-linear centroiding
scheme.  As we have seen, the photon weighted centroid is somewhat special
in that it is a {\sl linear\/} function of the atmospheric phase
fluctuation, and as a consequence, should have accurately Gaussian statistical
properties.  For non-linear centroiding or peak tracking the deflection
will not be strictly Gaussian --- the peak displacement will have 
discontinuities for instance --- but this does not seem to be a 
serious problem.  

There is another subtle difference between peak tracking and centroiding
which is how the evolutionary time-scale, and therefore the
necessary sampling rate, depends on the height distribution of
seeing.  If the wind speed is $v$ then the time-scale for
centroid motions is on the order of $\tau \sim D / v$, i.e.~just
the time it takes for the wind to cross the telescope pupil, regardless of whether
the seeing arises in a single screen or in multiple layers.
The speckle persistence time is also on the order $\tau \sim D / v$
for a single screen, but for multiple screens or a continuous
distribution the persistence time is predicted to be
$\tau \sim r_0 / v$ \cite{rgl82}.  This is rather worrying as it would suggest
that one would need much faster temporal sampling than the single screen
calculation suggests.  However, from numerical realizations  of
evolving PSFS (see \S\ref{sec:guidestars} below) 
we find this not to be the case; for $D / r_0 \simeq 4$
we find that the timescale for peak motions is very similar
for both single and multiple layer seeing, and that a sampling
rate $\simeq 0.5-1 D / v$ is adequate in either case.

\subsection{Pixelization Effects}
\label{subsec:pixelization}

In a regular CCD and with perfect guiding, the output image
is a set of point like samples of the 
convolution of the true sky with a box-like pixel function.
In an OTCCD device there is an additional degree of smoothing
because the image moves continuously, yet
the charge is shifted in discrete steps of finite size,
so there is some fluctuation in image position about the
effective pixel center which will have a box-like distribution
function.
In the design described in \S\ref{sec:detectors} below there are $\sim 10$
positions at which one can set the origin per physical pixel so this
extra smoothing is relatively minor, corresponding to a 10\%
increase in the pixel area.  To obtain a continuously sampled image
it is necessary to combine a number of exposures.  The image combination may
involve interpolation and this will introduce some further smoothing
of the image. The interpolation is however applied to the images after the photon counting
noise is realized, so there is no information loss in this step, and the
net effect of pixelization
on signal to noise is the same as applying a single convolution with the pixel function.

This sets a constraint on the pixel angular scale if the
pixelization is not to undo the improvement in image quality
provided by guiding. To quantify this we have taken the
exact fast guiding PSF and re-convolved with pixel functions of
various sizes and measured the reduction in Strehl ratio. 
We find that
a pixel size of $\pixelsize$ gives a reduction in Strehl of about
20\%, which we feel is just acceptable.  This sampling rate is about one half of the
critical sampling rate for this combination of telescope diameter and 
wavelength.

\subsection{Telescope Design Constraints}
\label{subsec:telescopes}

The main constraints on the design of the telescope are that it
should have a primary mirror diameter of about $\Dval$m and should be able to
give diffraction limited images over a square field of side 1 degree
or thereabouts.  A further constraint is imposed by the cost of
the detectors.  Since their cost scales roughly as the area of
silicon (rather than as the number of pixels) one would like to
make the pixels as small as is practical. As we discuss
below in \S\ref{sec:detectors} a pixel size of $\pixsizeinmicrons\mu$m
seems reasonable.  

In order to meet these requirements, 
we have explored several modified Ritchey-Chretien telescope designs
employing a refractive aspheric corrector located near the
focal plane.  The designs give diffraction limited images
over the required field of view,
and we have concluded that these designs are readily buildable
for a reasonable cost (see section 7). 

We can also consider all-reflective systems to avoid diffraction
spikes and 
scattered light from bright stars that may be a problem
with refractive correctors.
Such all reflective designs exist
and we expect that they could be implemented for a comparable
cost.

\section{Guide Star Constraints}
\label{sec:guidestars}

As discussed, in the Introduction, each telescope needs to measure
positions of hundreds of guide stars scattered
over the field. 
This is possible if the primary detector (a segmented OTCCD as
described in more detail in \S\ref{sec:detectors})
also serves as the guide star sensor.  This has the advantage of avoiding
the complication and expense of pick-off mirrors
and auxiliary detector, and by fast read-out of a 
small patch around each guide star one can sample
at rates in excess of 100~Hz which is more than adequate.
Disadvantages of this approach are that one then loses 
that element of the CCD array, of say an
arc-minute in size, for science, and that 
the guide stars must be observed through whatever filter is needed
for the science measurements, with concurrent loss of
photons.
The purpose of this section is to provide estimates of the
rate at which photons are collected as a function of telescope
aperture and guide star brightness, the rate at which we must sample
image motions, and the constraints these place on the number of
usable guide stars.

Scaling from the performance of existing thinned CCD cameras
on Mauna Kea, we expect that a
good, backside illuminated CCD will accumulate about one electron per second
from a source of $R$ magnitude
\begineq
m_1 = 24.6 + 5\log(D/1.5\hbox{\rm m}),
\endeq
(in the I-band the corresponding value is $24.3$).
Hence an exposure of $\Delta t$ of a source of magnitude $m$ should
yield $N$ electrons:
\begineq
N = \Delta t \times 10^{-0.4(m-m_1)}.
\endeq
This is the total number of electrons.  With fast guiding, what is more
relevant is the number of electrons in the diffraction limited core of
the PSF which is $N_{\rm core} = \alpha N$ with $\alpha \simeq 1/3$
for peak tracking.

The centroid or peak position will vary with time primarily because the
deflecting screen is being convected along at the wind speed.  In what follows we will
adopt a fiducial speed of $15$m/s.  This converts to a coherence
time for peak motions of $\tau \sim D / v = 0.1 {\rm s} (D / 1.5{\rm m})
(v / 15 {\rm m/s})$.
For given guide star brightness there is an optimal
choice of sampling rate, since if the sampling rate is too high then the
star centroid or peak location will be uncertain because of 
measurement error, while if the sampling rate is too slow the time averaged
position will not accurately track the instantaneous position.
To make this more quantitative we have made simulations in which
we generated a large Kolmogorov spectrum phase screen from which extracted
a sequence of closely spaced pupil-sized sub-samples, and for each of
which we computed the instantaneous PSF using (\ref{eq:otf0}).  
This stream of PSFs was averaged in pixels with appropriate 
angular size according to 
subsection \ref{subsec:pixelization} above, and averaged in time with
with some chosen integration time, and the result was then converted to
a photon count by sampling in a Poissonian manner and adding read noise,
the mean of which was taken to be $2e^-$. 
A simple peak-tracking algorithm --- 
locate the hottest pixel and then refine the position using 1st and
2nd derivative information computed from the neighboring pixels ---
was then applied to the simulated
pixellated
images, and the PSF for target objects was calculated by shifting the
instantaneous PSFs to track the peak and then averaging.
This calculation was performed for a coarse grid of star luminosities and
integration times.
\hide{
The results are shown in figure
\ref{fig:strehlvsintegrationtime} in which we plot the normalized
Strehl ratio for target objects as a function of integration time for a 
variety of guide star luminosities (expressed as the number of electrons per second). 
}
As expected, for bright objects we find the
optimal sampling rate is quite high, while for fainter objects the measurement
uncertainty tilts the balance towards longer integration times.
A good compromise for realistic guide stars is to take an integration
time of about $D / v$ corresponding to a sampling rate of about
10 Hz for our fiducial $15$m/s wind speed and diameter of
$D = 1.5$m. 
For very bright stars this is not optimal, but the
gain obtained by sampling faster is rather small.  With this sampling, we find 
negligible reduction in Strehl (as compared to rapid sampling of a
very bright guide star) for stars which generate about 4000 electrons per second,
or about 400$e^-$ per sample time, of which $\sim 140$ are in the diffraction
limited core. This corresponds to
an $R$ magnitude limit of
\begineq
m_{4000} = 15.6 + 7.5\log(D/1.5\hbox{\rm m})
\endeq
($15.3$ in the I-band).
The number of stars per square degree at the north galactic pole
brighter than $R$ magnitude $m_R$ is approximately 
\begineq
\label{eq:bscountsmodel}
\begin{matrix}{
N({<}m_R) = 2.8\times10^{-9} m^{9.2} \cr
N({<}m_I) = 5.6\times10^{-9} m^{9.1}
}\end{matrix}
\endeq
Equation (\ref{eq:bscountsmodel}) is a good fit to the \citeN{bs81}
model over the range $12 < m < 20$, and gives a number of sufficiently
bright guide stars per square degree of approximately
\begineq
N({<}m_{4000}) = 265\times(1+4.4\log(D/1.5\hbox{\rm m})
\endeq
($340$ in the I-band).

\begin{figure}[htbp!]
\centering\epsfig{file=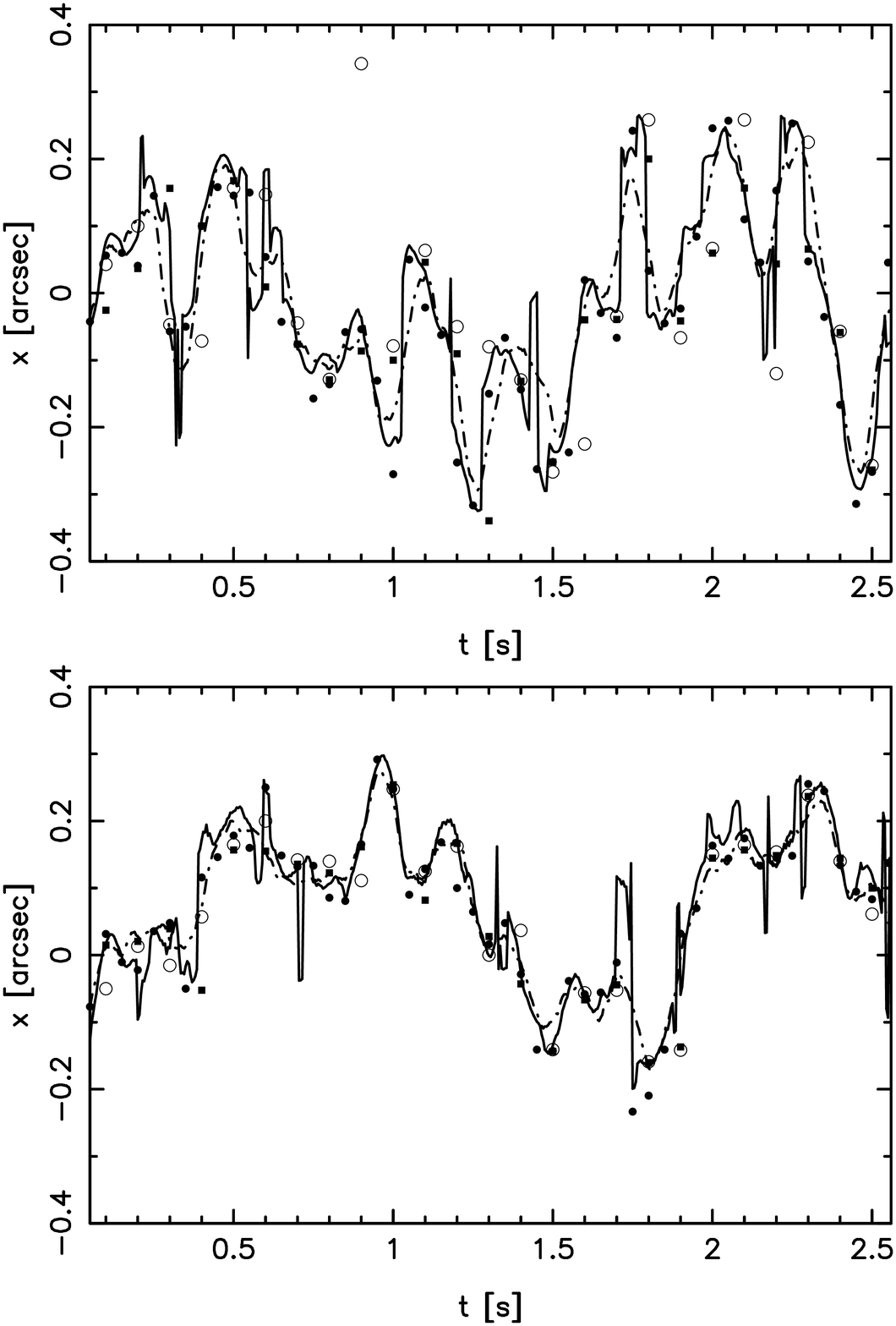,width=0.6 \linewidth}
\caption[Peak-tracking simulation.]
{The solid line shows one component of the deflection of the
peak of the PSF from a simulation performed as described in the
text.  Distance is converted to time here using an assumed wind speed
of $12.5$m/s.  The dot-dash line shows the trajectory of the
centroid.  The filled symbols correspond to a guide star of
magnitude $m_{4000}$ with a sampling rate of 20Hz (circles) and
10Hz (squares).  The open circles are for a star 4 times fainter and
sampled at 10Hz. The upper panel shows a realisation for a single
phase screen while in the lower panel four screens with the same speed but with
directions rotated though 0, 90, 180 and 270 degrees were used.
}
\label{fig:peaktracking}
\end{figure}

This number density corresponds to a mean separation $\simeq 3'.7$.
As this is greater than the coherence angle for
turbulence at altitudes higher than a few km, an instantaneous
measurement of the deflections of a set of guide stars does not
provide sufficient information to fully determine the deflection
field.  In making these estimates we have erred somewhat on the
side of caution.  For example, for stars which are a factor 4 
($\sim 1.5$ magnitudes) fainter than
the limiting magnitude quoted above the resulting target image Strehl is
reduced by about 30\%, so there is still a reasonable fraction of light in the
diffraction limited core (around $\sim 20$\% rather than $\sim 30$\% of the total)
and this increased the number density of stars by about a factor 2.5.
This result is illustrated in figure \ref{fig:peaktracking}.
Also, observations at lower galactic latitude will yield higher 
guide star densities by a factor ${\rm sec}(b)$, but the conclusion remains
that over most of the sky 
the sampling provided by natural guide stars is somewhat too
low to fully determine the deflection field from high altitude seeing.
In the next section we explore how one can obtain complete
sampling of the deflection --- and therefore guide out image motions
for all objects --- using an array of telescopes and using the past
history of guide star motions.

\section{Deflection Correlations and Guiding Algorithm}
\label{sec:correlations}

In this section we explore the correlations between neasured deflections of
guide stars and how to use these to compute the
deflection field needed to control the OTCCD charge shifting.
We first discuss the properties of conditional mean field estimators
which seem particularly appropriate for the problem.
We then consider the case of a single thin layer of
high altitude turbulence, and then discuss the generalization to
multiple or thick layers of turbulence, including ground level turbulence.

\subsection{The Conditional Mean Field Estimator}
\label{subsec:conddef}

The problem here is that we  wish to infer the deflection for a target object 
from measurements of the
deflection for a set of guide stars.  There are
various ways one might do this.  The approach we prefer is to
use the {\sl conditional mean deflection\/}.  

Consider first the simple case of
a 1-component Gaussian field $f(r)$ with correlation function $\xi(r) = \langle 
f(r') f(r' + r) \rangle$ and where we have a single measurement of $f$.  
The conditional probability distribution
for the field $f_1$ at some point $r_1$ given a measurement
of the field $f_2$ at some other point $r_2$ is
\begineq
p(f_1 | f_2) = p(f_1 , f_2) / p(f_2) 
\endeq
which is Bayes' theorem.  
According to the central limit theorem,
the joint probability distribution $p(f_1 , f_2)$
is given by (\ref{eq:multivariatepdf}).  Here the
covariance matrix is simply 
$M_{IJ} = \{\{\xi_0, \xi_r\}, \{\xi_r, \xi_0 \}\}$ so (\ref{eq:multivariatepdf}) yields
\begineq
\label{eq:conddef}
p(f_1 | f_2) = 
\sqrt{\xi_0 \over 2 \pi (\xi_0^2 - \xi_r^2)}
\exp\left(- \half {\xi_0 (f_1 - (\xi_r/ \xi_0) f_2)^2 \over 
\xi_0^2 - \xi_r^2}\right)
\endeq
This conditional PDF is just a shifted Gaussian with
conditional mean ${\overline f}_1 = (\xi_r / \xi_0) f_2$ and with
variance $\sigma^2 \equiv \langle (f_1 - {\overline f}_1)^2 \rangle = (\xi_0^2 - \xi_r^2) / \xi_0$.
At small separations the conditional mean field is equal to the
measured field, but relaxes to zero with increasing separation as $\xi_r / \xi_0$.
Thus the conditional mean estimator fails gracefully in the absence of useful
information (i.e.~far from the measurement point).  Compare this
with the behavior for an alternative, which is to use a maximum
likelihood estimator.  The likelihood is defined as the probability of the
`data', in our case $f_2$, as a function of the parameter $f_1$.
The likelihood is therefore
\begineq
L(f_1) = p(f_2 | f_1) = p(f_1 , f_2) / p(f_1)
\endeq
which is the same as the expression for the conditional probability, but with
$f_1$ and $f_2$ interchanged so from (\ref{eq:conddef}) it follows that 
the value of $f_1$ which maximizes the likelihood is
$f^{\rm ML}_1 = (\xi_0 / \xi_r) f_2$.  Like the conditional mean, this is
effectively
equal to the measured field if the separation $r_1 - r_2$ is much less than a
correlation length, so $\xi_r \rightarrow \xi_0$, 
but the solution blows up when the separation
becomes large and $\xi_r \rightarrow 0$, which is clearly undesirable for our application. 

A rather general feature illustrated by this simple example is that
the conditional probability also provides one with a measure of the
variance in the conditional mean field estimator 
$\sigma^2$, which is zero close to the
measurement
point and rises to become equal to the unconstrained variance $\xi_0$ at points
sufficiently distant from the measurement point that the correlation with the
measurement effectively vanishes.

The simple example of a single measurement of a 1-component field is readily 
generalized to the case where we have multiple measurements and
wish to constrain multiple target field values (the two components
of the target deflection for example).
Let us assume that one has $n$ target values $f_i$, $i = 0, n-1$ which one
would like to constrain with $m$ measurements $f_l$, $l = n, n + m -1$.
Let the covariance matrix be $M_{IJ} \equiv \langle f_I f_J \rangle$,
where $I$, $J$ range from $0$ to $N - 1$ with $N = n + m$.
The joint conditional mean probability distribution is
\begineq
p(f_0, f_1 \ldots f_{n-1} | f_n, f_{n+1} \ldots f_{n+m-1}) 
\propto p(f_0, f_1, \ldots  f_{n+m-1}) 
\propto \exp(- M_{IJ}^{-1} f_I f_J / 2)
\endeq
Ignoring factors which are independent of the target values 
$f_0 \ldots f_{n-1}$ this can be written as
\begineq
p( \ldots f_i \ldots | \ldots f_l \ldots)
\propto \exp(- m^{-1}_{ij} (f_i - \overline f_i) 
(f_j - \overline f_j) / 2)
\endeq
where $i$, $j$ range from $0$ to $n-1$ and repeated indices are
summed over.
This is a shifted Gaussian with conditional mean
\begineq
\label{eq:conditionaldef}
\overline f_i = m_{ij} \sum\limits_{I=n}^{N - 1} M^{-1}_{jI} f_I
\endeq
and with $n \times n$ error matrix
\begineq
\label{eq:errormatrix}
m_{ij} \equiv \langle (f_i - \overline f_i) (f_j - \overline f_j) \rangle 
= (M^{-1}_{ij})^{-1}.
\endeq
Note that the meaning of this equation is to take the upper-left
$n \times n$ sub-matrix from the inverse of the large matrix $M_{IJ}$ and
to invert this.  
Equation (\ref{eq:conditionaldef}) provides our
estimate of the target values, while
equation (\ref{eq:errormatrix}) provides the uncertainty in these
estimates.

\subsection{A Single Thin Turbulent Layer}
\label{subsec:singlethinlayer}

\begin{figure}[htbp!]
\centering\epsfig{file=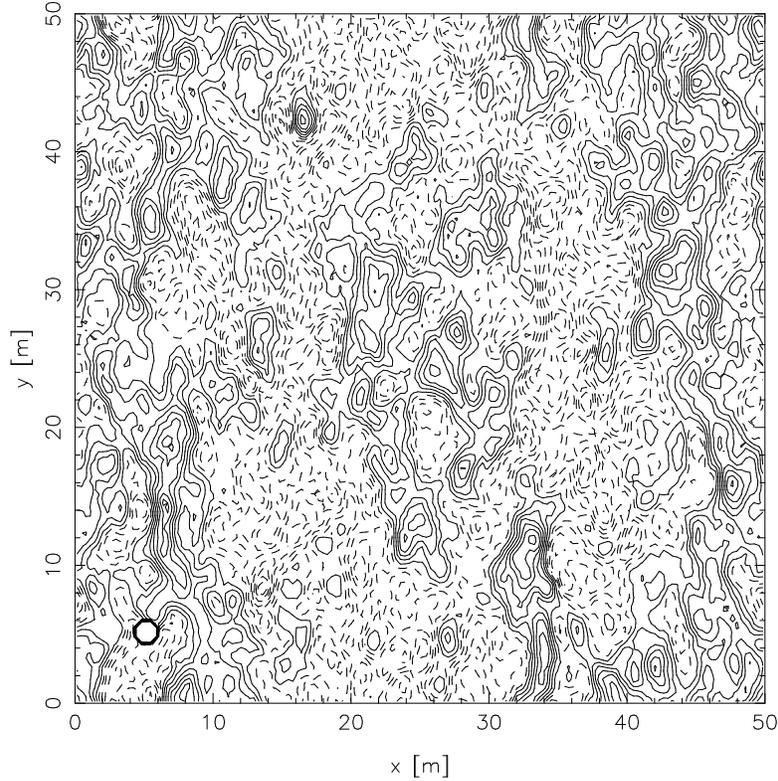,width=0.6 \linewidth}
\caption[Realization of deflection field]
{A realization of the deflection field $\delta \theta(\br)$ for
Kolmogorov turbulence.
This image was created by generating a white noise $P(k) \propto k^0$ Gaussian
random field; smoothing it with a power-law transfer function to create
an image with spectrum $P(k) \propto k^{-11/3}$ as expected for the
phase fluctuation for wavefronts passing though Kolmogorov
turbulence, taking the gradient (the $x$-component is
shown here), and finally convolving with
a telescope beam pattern.  The box side here is $50$m and the
telescope beam diameter is $1.5$m and is indicated by the disk
in the lower left. For a turbulence spectrum of
von Karman form with an outer scale of $20$m say, the
large-scale fluctuations would be reduced somewhat, but the general 
properties of the field are not much affected.
Note how the ridges and troughs of this function have a tendency to
be oriented vertically (i.e.~perpendicular to the component of the
deflection we have chosen to display).  This is a graphic illustration of
the property that transverse deflections have a greater range of
correlation than parallel deflections.
}
\label{fig:deflection}
\end{figure}

Source motions due to high altitude turbulence are expected to
be correlated only over limited angular separations.
The statistical character of the centroid deflection field
is illustrated in figure \ref{fig:deflection} which shows the deflection
field expected for a single turbulent layer. The patch shown here is
$50$m on a side and would subtend about $0^\circ .5$ at an
altitude of $5$km.  With typical wind velocities of say $15$m/s
this patch would be convected through its own length in a few seconds.

\begin{figure}[htbp!]
\centering\epsfig{file=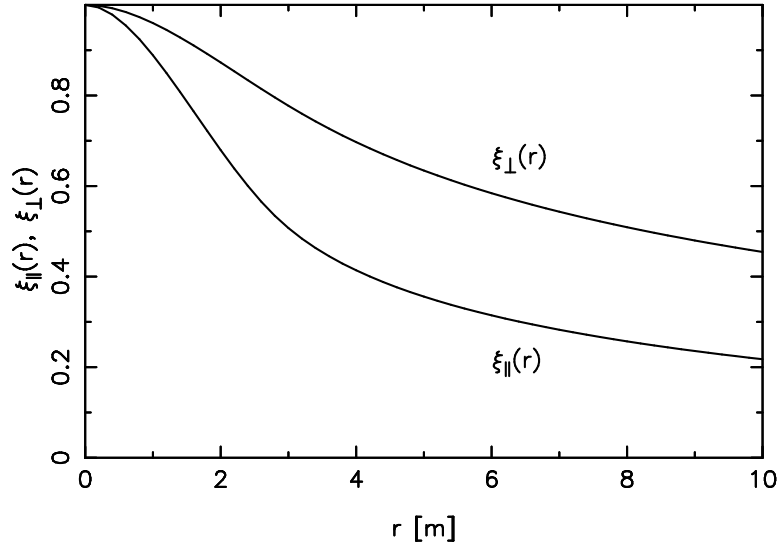,width=0.6 \linewidth}
\caption[Parallel and perpendicular transverse correlation function]
{These curves show the  parallel and transverse deflection correlation functions,
according to equations (\ref{eq:xidef})
assuming a telescope diameter of $1.5$m.
The parallel component of the deflection decoheres quite
rapidly with substantial decorrelation at lag of $\sim 1$m. 
Transverse deflections are correlated over greater range.
}
\label{fig:covariance}
\end{figure}

The range of correlations between deflections is 
shown more quantitatively in
figure \ref{fig:covariance}.  
Since the deflection is a vector, its covariance function
is a tensor: $\xi_{ij}(\br) = \langle \delta \theta_i(\br') \delta 
\theta_j(\br' + \br) \rangle$.
In a frame in which the lag $\br$ is parallel to the $x$-axis this is diagonal
and we define $\xi_\parallel(r) = \xi_{xx}(r)$ and $\xi_\perp(r) = \xi_{yy}(r)$.
One can then obtain $\xi_{ij}(\br)$ in the general frame by applying the rotation
operator.  The parallel and perpendicular deflection correlation functions
are given by
\begineq
\label{eq:xidef}
\begin{matrix}{
\xi_\parallel(r) = \int {d^2k \over (2 \pi)^2} k^2 P(k) W^2(k) (J_0(kr) - J_2(kr)) \cr
\xi_\perp(r) = \int {d^2k \over (2 \pi)^2} k^2 P(k) W^2(k) (J_0(kr) + J_2(kr))
}\end{matrix}
\endeq
where $J_0$, $J_2$ are Bessel functions, and
with
\begineq
W(k) = \int\limits_0^{D/2} dr \; r J_0(kr)
\endeq
and where $P(k) \propto k^{-11/3}$ for Kolmogorov turbulence.
While one should really use diffraction
theory to compute how the image quality degrades with imperfect
guiding information, to an approximation which should be sufficiently
accurate for our present purposes we will adopt the `Friedian'
approximation and assume that the real PSF will be the
PSF for perfect guiding convolved with the distribution of
errors in the centroid estimate which can be inferred from
figure \ref{fig:covariance}.

In standard tip/tilt implementations the whole image is shifted,
with the shift determined from a single guide star.  According to
figure \ref{fig:covariance} this will give good image quality
within the angular scale which projects to one telescope beam width at the
altitude of the deflecting layer, or about $1'$ or so.
At larger separations the image quality will deteriorate, with
a tendency for the PSF to first become elongated in the radial direction
(because the radial component of the deflections decoheres more 
rapidly with distance),
and at very large separations the centroid motion will 
be totally uncorrelated with the motion of the distant guide star and 
this `over-guiding' will actually
cause a deterioration of the image quality as shown in figures 
\ref{fig:isoplanatism1}, \ref{fig:isoplanatism2}.
With an OTCCD array one can guide separate parts of the focal plane 
independently, and this opens up many possibilities for improvement.
Even with a single guide star, one can do better by using a guiding
signal which is the conditional mean deflection at the
point in question, given the measured deflection of the star
at some other point.  As discussed, the conditional mean will relax towards
zero at large distance from the guide star, and this will at least solve the
over-guiding problem.  

\begin{figure}[htbp!]
\centering\epsfig{file=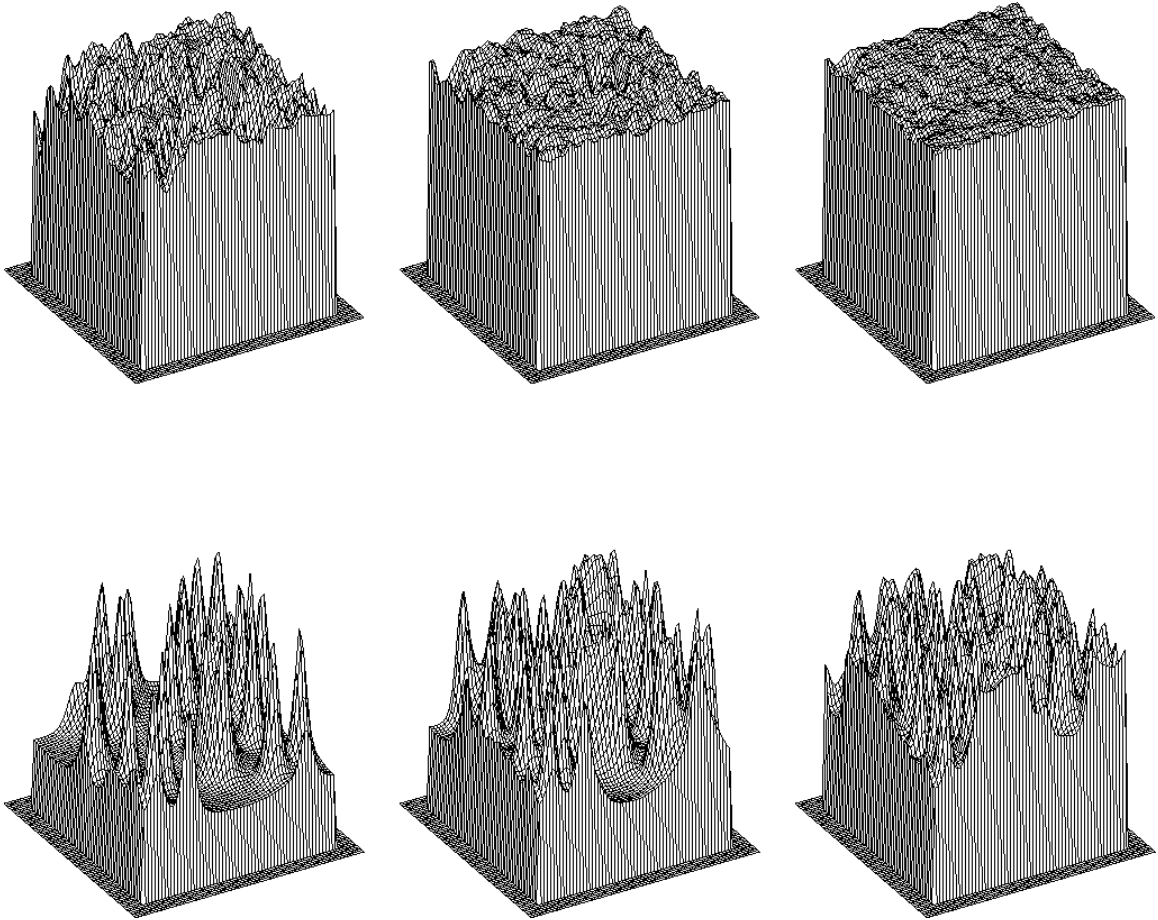,width=\figwidth}
\caption[Strehl distribution for multiple telescopes]
{The height of these surfaces represents the image quality
on a scale of $0-4\times$ the uncorrected quality
as a function of position across the field for various
number of telescopes.  Each panel shows a region of sky about
1/4 degree on a side, and a telescope diameter $D = 1.5$m and 
altitude $h = 5$km  were assumed.  
Lower left panel is for a single telescope and the other
panels show the results of increasing the sampling density by
a factor $N_T = 2,4,8,16$ and $32$. 
The measure of image quality used is the normalized Strehl ratio,
which is roughly proportional to the fraction of light contained
within the diffraction limited PSF core. This was computed from the
variance of the  conditional mean
centroid shift estimator.  Modeling the effect of uncertainty
in the predicted centroid as a convolution with a Gaussian
ellipsoid we have found that the normalized Strehl is given
approximately by Strehl $= 3 \times \exp(-\alpha / 0.18) + 1$ where
$\alpha = \sigma^2 / \sigma_{\rm total}^2$ is the fractional
variance in the conditional mean, with $\sigma_{\rm total}^2$ is
the unconstrained variance.  The normalized Strehl ratio 
saturates at $S \simeq 4$ when the uncertainty in the
centroid becomes small.
To compute these images, for each pixel we identified the nearest
$N \simeq 10$ stars, and used the theoretical deflection
correlation functions to compute the $(2 N + 2) \times (2 N + 2)$ covariance
matrix relating the 2 deflection components of the target
point to those of the neighboring stars, and then inverted this
to obtain the covariance matrix for the conditional mean.
}
\label{fig:conddef}
\end{figure}

More interestingly, we can use the
multi-variate conditional probability machinery to compute a conditional
mean deflection field given measurements of a number of guide stars.
Unfortunately, with a single telescope,
most target points are too far from guide
stars to gain much improvement.
This is illustrated in the lower left panel figure \ref{fig:conddef} which shows the
variation in image quality, which was computed from the uncertainty in the
conditional mean deflection, given a
set of measurements of the deflection --- assumed to be due to a single layer of
turbulence ---
for a set of randomly placed guide stars with realistic number density.
Aside from generally small islands of small error very close to the
guide stars, there is large uncertainty in the centroid motion and
the most probable deflection will be quite small in these uncertain regions
and the increase in image quality quite meager.
More complete sky coverage is possible with an array of telescopes.
Adding extra telescopes which monitor the motions of the same set of
guide stars will provide additional samples of the deflection field with the
same pattern as for a single telescope, but stepped across the 
deflecting screen by the
telescope separation, as was illustrated in figure \ref{fig:beamsplot}.  
If the telescope spacing is much greater than the
mean separation of guide stars then the result
is essentially a Poisson sample of deflections with sampling density
enhanced by a factor $N_T$, this being the number of telescopes in the array.
The results for various $N_T$ values
are shown in figure \ref{fig:conddef}.
The image quality increases 
dramatically with the number of telescopes.  With $N_T = 16$,
the typical fractional position variance is $\sim 10^{-2}$, which 
is very accurate indeed,
and essentially all points on the sky have image quality close to the
maximum allowed, so with this sampling
density one can accurately predict the motion of any target object.
Note that as the sampling density is increased there is a rather sharp
transition as the area wherein the deflection is well determined `percolates'
across the field.

Figure \ref{fig:conddef} can also be interpreted as giving the
performance of a single telescope for a single deflecting layer at
altitude $h = 5 N_T^{-1/2}$km.
Thus, under favorable conditions one could expect to obtain
good performance with a single instrument, but this would be the
exception rather than the rule.

\begin{figure}[htbp!]
\centering\epsfig{file=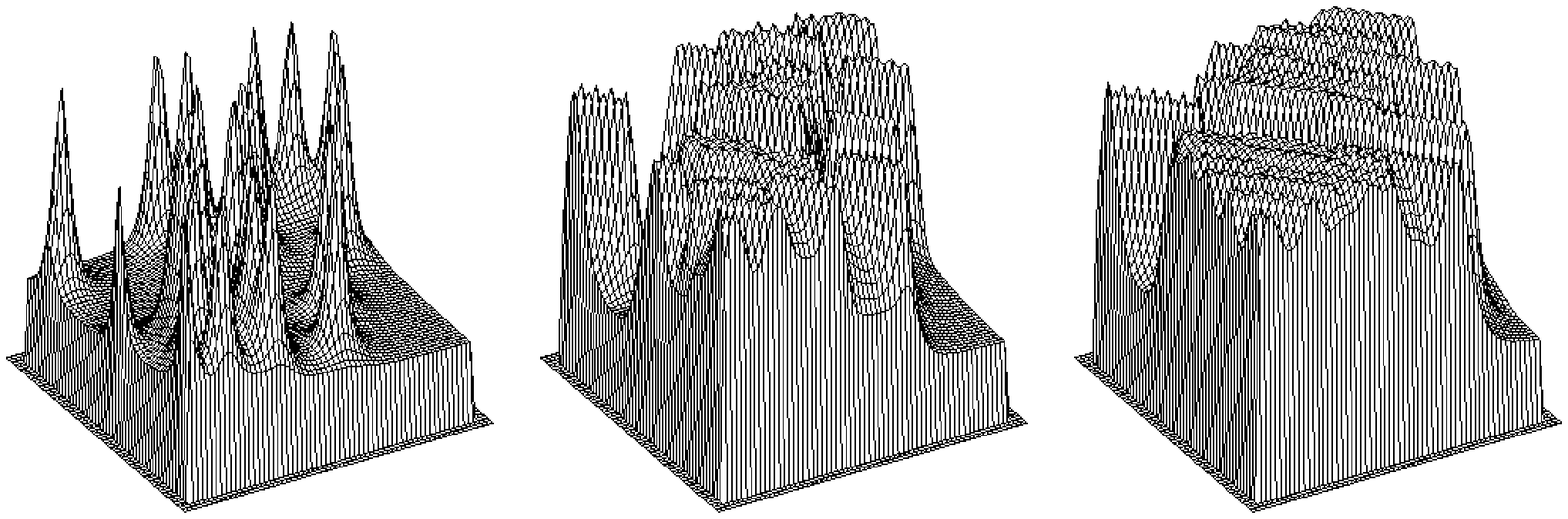,width=\figwidth}
\caption[Strehl distribution for a single telescope using time history.]
{These surfaces show the variation of image quality for a single
telescope, but using the time history of the deflection field
under the `frozen-turbulence' assumption. 
Vertical scale is as in figure \ref{fig:conddef}. Left hand panel is
for a single measurement, while center and right panels show the
effect of using 5 and 10 measurements spaced by about 1m in
physical separation, corresponding to a sampling rate of
$10 (v/10{\rm m/s}) {\rm Hz}$.  
}
\label{fig:conddefx}
\end{figure}

In this analysis only the instantaneous guide star deflections
were used.  Under the `frozen turbulence' or `Taylor flow'
assumption there is more information at our disposal 
encoded in the history of the
guide-star deflections, which 
provide a set of line-like samples of the deflection fields
lying parallel to the wind direction.
For a given
target point, the most valuable information will come from those
guide-stars lying
up-wind and at a time lag given by the spatial separation
divided by the wind speed. In the frozen turbulence assumption
the mean number of such trails
passing within say  $\pm 1'$ of a given point is roughly
$n \Theta N_T$, where $n$ is the number density of guide stars,
and $\Theta$ is the field size (using angular units of arc-minutes).  
For a field size of one degree and
$n = (4')^{-2}$ say, this we expect about $4 N_T$ trails on average
passing within a correlation length of a typical target point.
For $N_T \sim 30$ this would give a very
dense sampling rate.  However, it
may be over-optimistic, as it assumes that the turnover
time for $\sim D$ sized eddies is as long as the time-scale for
the layer to sweep across the whole field $\sim h \Theta / v$ which is
several seconds.  If the turnover time
is shorter, then the correlations will decay more rapidly, and
one should replace $\Theta$ in the sampling density estimate 
by the angular distance an eddy
propagates in one turnover time.  Anecdotal evidence 
suggests that the
frozen turbulence approximation appears to be well obeyed 
over scales of several meters,
so this would
suggest that one should expect to obtain a substantial gain in sampling
density 
by using temporal information.  The improvement of image quality
afforded by temporal history is shown in figure \ref{fig:conddefx}
assuming that the turbulence is effectively frozen for
displacements of 5m and 10m respectively, corresponding to 
times of $0.5-1$s for wind speed of 10m/s.
This is quite promising as it shows that even with a single
telescope one can obtain good image quality over large areas of the
sky, and that with just a few telescopes one should be
able to obtain essentially full coverage.
We caution, however, that this conclusion is strongly dependent on
the assumption of a single deflecting screen.

In the calculations shown in figures \ref{fig:conddef}, \ref{fig:conddefx} 
the height of the
deflecting layer was assumed known and the covariance matrix $M_{IJ}$ 
was computed from (\ref{eq:xidef}).  In reality we would need to 
measure the covariance matrix from the actual measured guide star
deflections.  For a thin deflecting layer at height $h$ and moving
with velocity $\bv$ the deflection covariance function is 
\begineq
\langle \delta \theta_i(\bx, \btheta, t) \delta \theta_j(\bx', \btheta', t') \rangle
= c \xi_{ij} (\Delta\bx + h \Delta\btheta + \bv \Delta t)
\endeq
where $\delta \btheta(\bx, \btheta, t)$ is the deflection
of a guide star with position $\btheta$ on the sky
seen with a telescope at position $\bx$ at time $t$, $\Delta \bx 
\equiv \bx - \bx'$ etc.~and $c$ is a measure of the intensity of the layer.  
The covariance function can be estimated by averaging products of pairs
of deflections.  For a regular grid array of telescopes and uniform
sampling in time, one obtains samples of the covariance on a uniform
grid in $\Delta\bx$, $\Delta t$ space, which is just what one needs.  
The sampling in angle space $\Delta \btheta$ is a little more problematic
since the guide stars are randomly distributed, and yet one needs
to evaluate the covariance at the angular separation between a guide star and
the target object, which will not in general coincide with the separation
between any particular pair of guide stars.  To solve this one can
bin the pairs into a grid of finite cells in $\Delta \btheta$ space
and, if necessary, interpolate to obtain the covariance at the desired 
separation.  This should not be too difficult.
If one has say $N_s \sim 200$ guide stars on a $\Theta = 1^\circ$ field
then the number of pairs per $\sim (1')^2$ correlation area is $\sim 5$
which should be adequate; the probability of having an empty cell if we bin
into cells of $\sim 1'$ size is very small, and interpolation over any
patches should be fairly safe.  

In principle, one could compute the covariance matrix for all pairs of observations
and then invert the resulting matrix. Computing the full covariance matrix
would be time consuming, but not insurmountably so.
For $N_s = 200$ guide stars, and $N_T = \Ntval$ telescopes,
and if we keep a running history of say $N_t = \Ntauval$ previous measurements then
we have $(2 N_s N_T N_t)^2 / 2 \sim 10^{11}$ pairs of measurements at any one time.
Since the time history will be uniformly sampled the
$\Delta t$ correlations can be performed with a FFT, and similarly 
for the $\Delta x$ correlations if the
telescopes are laid out on a regular grid.  With this simplification, the
time complexity of the covariance matrix accumulation is essentially
that of performing $N_s^2 / 2$ small ($N_t \times N_T$) 
FFTs every second or so which is not overwhelming since commercially available
DSP devices perform FFTs at a rate of $\sim 50$M floating point
data values per second.
The real problem with this `sledgehammer' approach is that to compute the
conditional mean we will need to {\sl invert\/} this huge 
$(2 \times N_T \times N_s \times N_t)^2$ matrix, which is
prohibitively expensive in computational effort and is probably also numerically
unstable.  Luckily we do not need to do this.  As discussed, the most valuable
information pertaining to the deflection of a target object will come from the
relatively small number of guide stars that are seen through the same
patch of turbulence at some up-stream position.  It is easy to identify which
these observations are since they are those for which the correlation 
with the target object deflection is particularly
strong, and so one need only work with the small subset of the
full covariance matrix that involve these critical observations, and 
this greatly reduces the amount of computation.  
This was the approach used in computing figure \ref{fig:conddef} where only
a subset of the guide stars were used for each pixel of the image.

The matrix
inversion need only be done infrequently; on the rather long time-scale for
macroscopic conditions such as wind speed, turbulence strength etc.~to 
change. An instantaneous measurement of
the covariance function will of course be noisy as we are sampling
a single realization of a random process with finite extent in
size and time.  However, since the correlation time is on the
order of the eddy turnover time of perhaps a few seconds, 
we can obtain statistically independent
samples at intervals of order this time, so by averaging over
several minutes say one should be able to determine the 
ensemble average covariance very accurately. 
The computation of the mean deflection needs to be done on
a very short time-scale (a few tens of Hertz) but this is computationally
a much easier task.
A pleasant feature of this approach is that one can be fairly liberal in selection
of guide stars; faint stars give more uncertain positions which
therefore correlate less with other more precisely measured  motions. The correlation
matrix machinery `knows' this and will automatically give these stars the
weight they deserve.

Equation (\ref{eq:conditionaldef}) provides the guide signal for
the target cell of the detector.  As discussed, in the Friedian
approximation, the actual PSF is the PSF for perfect
fast guiding convolved with a Gaussian ellipsoid
$\exp( - \bx \cdot \bm^{-1} \cdot \bx / 2)$.  This is
useful since the image quality for a given patch of sky will
vary from telescope to telescope and with position within the field, so
(\ref{eq:errormatrix}) provides a useful criterion for 
rejecting or down-weighting poor images or sections thereof.

The procedure outlined above is somewhat inefficient in that it
requires the computation of a fairly large matrix 
for each target.  Neighboring target cells will tend to
correlate strongly with the same set of guide stars, so the
set of stars which correlate strongly with one or
more of a cluster of neighboring cells will likely be not
much larger than the set which correlate strongly with
any individual cell. If so, then a great saving in computational
effort can be made by computing the conditional probability for
the deflections for the cluster of target cells at one go.

\subsection{Multiple or Thick Turbulent Layers}
\label{subsec:multilayer}

The foregoing analysis was somewhat idealized in that a single thin layer of
turbulence was assumed.  If there are multiple or thick layers then the situation
is somewhat more complicated.  
Nonetheless, given the huge amount of information at our disposal
we believe that the conditional 
probability approach should still work, though depending on the nature of
the turbulence there may be non-trivial constraints on the layout of the
telescope array.

Consider first the case of a single {\sl thick} layer at high altitude. The
procedure outlined in the previous section will then fail
if the baselines between the telescopes are too large.  The problem is that
the deflection for a target object at the zenith say will 
sample a vertical tube through 
the layer while a guide star seen from a different telescope
at distance $\Delta \bx$ will sample a tube which is inclined at an
angle $\sim \Delta \bx / h = \Delta \btheta$, and even if these two tubes overlap
exactly in the center of the layer the deflections will tend to decohere if
the thickness of the layer exceeds the value $\delta h \sim D h / \Delta x$.
Unfortunately, the observational situation is somewhat unclear as e.g.~SCIDAR
measurements tend to be limited in height resolution.  For a width $\delta h \sim
100$m say, and $h \sim 5$km this implies the constraint on the
size of the array $L \lsim D h / \delta h \simeq 50 D \sim 75$m. 
This is not very restrictive. 
A further possible complication is wind shear across a thick layer
which will tend to modify the deflection correlations.

Now consider multiple deflecting layers.
According to the admittedly rather patchy observational studies reviewed
above, one quite common situation is for there to be an additional 
strong component of seeing coming from the planetary boundary layer and from
the immediate environment of the telescope, the so-called `dome seeing'.
This is quite easy to deal with since it causes a common deflection
for all of the guide stars for each telescope.  If we simply take the mean
deflection and subtract this, provided we have numerous guide stars and
a sufficiently wide field then the residual guide star deflections
after we subtract the mean should be essentially those due to the high
altitude turbulent layer alone, and we can proceed as before.
There are other means at ones disposal to further reduce the
effect of low level turbulence. Very low level refractive index 
fluctuations can be homogenized by means of louvred enclosures
and/or fans. The telescopes here are light and are auto-guiding, so it is not
unreasonable to consider some kind of elevated support to raise them
above very low level seeing.
Also, since the isoplanatic angle for low level seeing is
very large one can consider doing higher order wavefront correction
with a deformable secondary.  One way to implement 
this would be to augment the individual telescopes with smaller 
telescopes deployed around the
periphery of the primary mirror which measure the average deflection
of say a few tens of bright guide stars within the field.  By taking the
average, one effectively isolates the effect of the dome and boundary
layer, and one can show that with say 6 such auxiliary telescopes --- which
provide an extra 12 constraints in the form of samples the peripheral
wavefront tilt --- one can effectively negate the effect of even quite strong low-level
seeing. Provided low level seeing can be effectively eliminated by one or
other of these approaches, 
one would then tune the aperture of the
main telescopes such that they have diameter  4 times the $r_0$ for the
`free-atmosphere' seeing alone, as we have assumed above.
 
More difficult is the case where there are two or more 
high altitude layers giving a significant 
and comparable contribution to the deflection. 
It would be tempting to argue that since
the strength of individual layers appears to have a highly
non-Gaussian distribution with large dynamic range, having several layers of 
very similar strength is statistically improbable.
However, this is probably over-complacent for the
following reason: In the scheme outlined above, and with a relatively
weak secondary deflecting layer, the correlation machinery will `lock
on' to the primary layer, with the net result that the 
sharp corrected image will get convolved with the natural
seeing PSF for the second level.  This can result in a significant loss
of image quality even for a rather weak secondary layer.  As a specific example,
a secondary layer contributing only 4\% of the total deflection
power produces a reduction in Strehl of about 25\%, so it is clearly
highly desirable to have a guiding algorithm which can
cope with multiple layers.

The problem here is not lack of information; with tens of telescopes, hundreds
of guide stars and many time-steps in the history of the deflections one has a
huge amount of information with which one should be able to separate the
effects of multiple layers.  In principle one could simply
compute and invert the full covariance matrix to obtain the conditional deflection.
Indeed, a rather nice feature of this sledgehammer approach is that there
is then no need to try to solve for a set of discrete layer heights and
intensities; the probability engine takes care of it automatically, as all the relevant
information is encoded in the covariance matrix that one measures.
The real problem here is how to decipher the
information in a realistic amount of time.  For a single layer it is relatively
easy to identify and isolate a relatively small number of critical measurements
which have a bearing on a given target object deflection. For multiple layers
this is not the case, and a different approach is required.

What is needed is some way to at least approximately diagonalize the covariance 
matrix, so that one can avoid having to invert it all at once.  
Given the
statistical translational invariance of the Gaussian random deflection screens
we are dealing with,
a natural approach is to work in Fourier space.
Let us assume that we have a set of discrete deflecting layers at heights $h$,
streaming across the field with velocities $\bv_h$, and let us
model the deflection field (which we will denote by ${\bf f}$ here) 
as a function
of telescope position $\bx$, angular position $\btheta$ and time $t$ as
\begineq
\label{eq:multilayermodel}
{\bf f}(\bx, \btheta, t) = 
\sum\limits_h {\bf f}_h(\bx + h \btheta + \bv_h t, t).
\endeq
In the perfect frozen flow limit ${\bf f}_h$ would depend on time only
implicitly through 
the spatial coordinate $\bx + h \btheta + \bv_h t$.
The inclusion of an
explicit dependence of ${\bf f}_h$ on time $t$ here allows for the
evolution of the deflection field due to the turnover of eddies, though 
as discussed, we
expect that the explicit time dependence will be rather
slow as compared to the typical induced time dependence due to the
motion of the screen.
The deflection field is a random function of position and time
and can be expressed as a Fourier synthesis
\begineq
\label{eq:fouriersynthesis}
{\bf f}_h(\bx, t) = \int {d^2 k \over (2 \pi)^2} {d \omega \over 2 \pi}
\tilde {\bf f}_h(\bk, \omega) e^{-i(\bk \cdot \bx + \omega t)}
\endeq
where the statistical homogeneity, stationarity and isotropy of the
individual phase screens and their assumed mutual independence implies that
distinct Fourier modes are statistically independent:
\begineq
\label{eq:powerspectrumdefinition}
\langle \tilde f_{hi}(\bk, \omega) \tilde f^*_{h'j}(\bk', \omega')\rangle = 
(2 \pi)^3 \delta_{hh'} \delta_{ij} \delta(\bk - \bk') \delta(\omega - \omega')
P_h(k, \omega)
\endeq
where $P_h(k, \omega)$ is the spatio-temporal power spectrum of the layer
at height $h$, and reality of ${\bf f}_h(\bx, t)$ imposes the symmetry
$\tilde {\bf f}_{h}(-\bk, -\omega) = \tilde {\bf f}_{h}^*(\bk, \omega)$.
For Kolmogorov turbulence the spatial power spectrum 
$P_h(k) \equiv \int d \omega \; P_h(k, \omega)$ is
a power law
$P_h(k) \propto k^{-5/3}$ at low spatial frequencies, with the smoothing
with the telescope pupil introducing a high-$k$ cut-off at 
$k \sim 1 / D$.  Kolmogorov theory also tells us that the typical
eddy velocity scales as the $1/3$ power of the eddy size, or equivalently
as $v \propto k^{-1/3}$, so the width of $P_h(k, \omega)$ in 
temporal frequency must scale as
$\omega_*(k) \sim v / L \propto k^{2/3}$.  An acceptable model for $P_h(k, \omega)$
for a thin layer of turbulence is then
\begineq
\label{eq:pkwmodel}
P_h(k, \omega) = P_h\Gamma(\bk, \omega) 
\quad\quad {\rm with} \quad \quad 
\Gamma(\bk, \omega) = k^{-5/3} \tilde T^2(k)
\zeta(\omega / \omega_*) / \omega_*(k)
\endeq
with $\omega_*(k) = \omega_D (k D)^{2/3}$ and where $\tilde T(k)^2$ is the
Fourier transform of the telescope pupil, which, for a simple
filled circular aperture, is the `Airy-disk' function.  
This model is parameterized by
an overall intensity $P_h$ and by $\omega_D$ which is
the turnover time for eddies of spatial frequency $k \sim 1 / D$.
The function $\zeta(y)$ is dimensionless, bell-shaped, and of unit width,
which we take here to be approximately Gaussian. 
In 3-dimensional $\bk - \omega$ space the model (\ref{eq:pkwmodel}) 
is a disk of with axis parallel to the $\omega$-axis,
radius $k_* \sim 1/ D$, and with scale height $\omega_*(k)$.
The assumed Gaussian form of the vertical profile $\zeta$ here is crude guess,
and one would want to modify this in the light of either empirical
observations and/or more sophisticated theoretical modeling.

Consider a particular telescope at position $\bx_0$ and
at the present time, which we take to be $t = 0$, and define the angular transform
of the deflection at angular frequency $\bkappa$ 
as ${\bf f}_0(\bkappa) \equiv \int_\Theta d^2 \theta \; 
{\bf f}(\bx_0, \btheta, t = 0) e^{i\bkappa \cdot \btheta}$,
where the subscript $\Theta$ on the integral indicates that the
integration is taken over the field of size $\Theta$.
Using (\ref{eq:multilayermodel}), (\ref{eq:fouriersynthesis})
we can express ${\bf f}_0(\bkappa)$ in terms of the
spatio-temporal Fourier mode coefficients as
\begineq
\label{eq:f0}
{\bf f}_0(\bkappa)
= \sum\limits_h 
\int {d^2 k \over (2 \pi)^2} {d \omega \over 2 \pi}
\tilde {\bf f}_h(\bk, \omega) e^{-i\bk \cdot \bx_0}
\overline W_\theta(\bkappa - h \bk)
\endeq
where 
$\overline W_\theta(\bkappa) \equiv 
\int_\Theta d^2 \theta \; e^{i \bkappa \cdot \btheta}$.
If we take the field to be a square of side $\Theta$ then this
is a 2-dimensional sinc function:
\begineq
\label{eq:Wbar}
\overline W_\theta(\bkappa)
=  \Theta^2 
{\sin^2 \Theta \kappa_x / 2 \over (\Theta \kappa_x / 2)^2}
{\sin^2 \Theta \kappa_y / 2 \over (\Theta \kappa_y / 2)^2}
\endeq
This function has a `central lobe' at $\bkappa = 0$ of height $\Theta^2$ 
and width $\delta \kappa \sim 1 / \Theta$ flanked by side lobes 
of oscillating sign which diminish
rapidly with increasing $\bkappa$,
so ${\bf f}_0(\bkappa)$ depends only on those Fourier modes
in a small range of spatial frequency 
of size $\delta k \sim 1 / h \Theta$ around $\bk = h \bkappa$, this being the
spatial frequency which maps to the chosen angular frequency $\bkappa$ at
altitude $h$.  If $h \Theta > L$ then the complex phase factor 
$e^{-i\bk \cdot \bx_0}$ in (\ref{eq:f0}) is nearly constant and
so ${\bf f}_0(\bkappa)$ is the product of $\tilde f(\bk, \omega)$
with a cylindrical window function
which is infinitely long in $\omega$ and
has width $\delta k \sim 1 / h \Theta$.  If on the other hand $h \Theta < L$
then the variation of the phase factor is appreciable for typical $\bx_0 \sim L$
and the window function has oscillatory modulation with scale
length $\sim 1 / L$.

Similarly, if we 
define the 5-dimensional Fourier transform of the measured deflections as
\begineq
\label{eq:Fdef}
\bF(\bk, \bkappa, \omega) = 
\sum\limits_{\bx, \btheta, t} {\bf f}(\bx, \btheta, t) 
e^{i (\bk \cdot \bx + \bkappa \cdot \btheta + \omega t)}
\endeq
then, in terms of the transform of the deflection screens 
$\tilde {\bf f}_{h}(\bk, \omega)$, this is
\begineq
\label{eq:Ffromtildef}
\bF(\bk, \bkappa, \omega) =
\sum\limits_{h'} \int {d^2 k'\over (2 \pi)^2} {d \omega' \over 2 \pi}
\tilde {\bf f}_{h}(\bk', \omega')
W_x(\bk - \bk') W_\theta(\bkappa - h \bk') 
W_t(\omega - \omega' - \bk' \cdot \bv_{h'})
\endeq
where $W_x(\bk) \equiv \sum_\bx e^{i \bk \cdot \bx}$ is the
Fourier transform of the telescope array;
$W_\theta(\bkappa) \equiv \sum_\theta e^{i \bkappa \cdot \btheta}$
is the transform of the guide star distribution and 
$W_t(\omega) = \sum_t e^{i \omega t}$ is the transform of the
of the temporal sampling pattern. 
Now all of these functions are quite strongly peaked at zero argument, $W_x$ has width
$\delta k$ 
of order the inverse of the telescope array size $L$,
$W_\theta$ has width $\delta \kappa \sim 1 / \Theta$ (like $\overline W_\theta$)
and $W_t$ has width $\delta \omega$ 
equal to the inverse of the time period $T$ over which
we choose to integrate.  Consequently, and like ${\bf f}_0(\bkappa)$,
$\bF(\bk, \bkappa, \omega)$ 
receives a large contribution from a restricted region of
spatial frequency around $\bk' \simeq \bk$.  
Unlike  ${\bf f}_0(\bkappa)$ however, the contribution to
$\bF(\bk, \bkappa, \omega)$ is also restricted in
altitude and temporal frequency since for the argument of 
$W_\theta$ to be small requires both that $\bkappa$ point in approximately the
same direction as $\bk$ and that the ratio of angular to spatial frequency
$\kappa / k$ should coincide with the
altitude of an actual layer of turbulence.
Finally, $\bF(h, \bk, \omega)$ is most
sensitive to temporal frequencies of the deflection 
screen $\omega' \simeq \omega - \bk \cdot \bv_h$ which means,
if we assume that the 
intrinsic deflection screen evolution
time-scale is long compared to $D / v$, that $\bF(\bk, \bkappa, \omega)$ is 
only sensitive to
a screen at altitude $h$ if the screen velocity 
satisfies $| \bk \cdot \bv_h - \omega | \le \omega_*$, where $\omega_*$
is on the order of the inverse of the eddy turnover time.  

Equations (\ref{eq:f0}) and (\ref{eq:Ffromtildef}) above give 
${\bf f}_0(\kappa)$ and
$\bF(\bk, \bkappa, \omega)$ respectively as integrals of $\tilde{\bf f}_h(\bk, \omega)$
times some window function.  However, the window
function for $\bF(\bk, \bkappa, \omega)$ is in all dimensions at least as narrow
as the window function for ${\bf f}_0(\kappa)$, and 
therefore ${\bf f}_0(\bkappa)$, from
which we can trivially extract the desired guide signal 
${\bf f}(\bx_0, \btheta, t = 0)$ by inverse
transforming, should be
well constrained by measurements of $\bF(\bk, \bkappa, \omega)$
taken at appropriate spatial, angular and temporal frequencies.
Since these are both linear functions of $\tilde {\bf f}(\bk, \omega)$
they should have Gaussian statistics, and so one can write down the
conditional probability
\begineq
\label{eq:poffk}
p({\bf f}_0(\bkappa) | 
\bF(\bk', \bkappa', \omega'), \bF(\bk'', \bkappa'', \omega'') \ldots)
\endeq
from which one can determine the conditional mean value of 
${\bf f}_0(\bkappa)$, very much as was done before for a single deflecting layer.
Let us assume for the moment that $P_h(\bk, \omega)$ and the
velocities $\bv_h$ are known.  If so, the covariance matrix
in the multi-variate Gaussian distribution (\ref{eq:poffk}) has components 
like
\begineq
\begin{matrix}{
\langle f^*_{0i}(\bkappa') F_j(\bk, \bkappa, \omega) \rangle = \cr
\delta_{ij} 
\sum\limits_h \int {d^2 k'\over (2 \pi)^2} {d \omega' \over 2 \pi}
P_h(k', \omega')  
e^{-i\bk' \cdot \bx_0} 
W_x(\bk - \bk') W_\theta(\bkappa - h \bk') 
W_t(\omega - \omega' - \bk' \cdot \bv_h)
\overline W_\theta^*(\bkappa' - h \bk')
}\end{matrix}
\endeq
Since the various window functions here are known and compact, it is straightforward
to enumerate the limited number of 5-dimensional transform modes which
are relevant for any given $\bkappa$, evaluate the appropriate
covariance matrix and thereby obtain an accurate conditional mean
estimator for ${\bf f}_0(\bkappa)$ 
as a linear combination of a limited subset of the $\bF(\bk, \bkappa, \omega)$ values.

To implement this program we need some way of determining $P_h(k, \omega)$.
If we evaluate $\bF(\bk, \bkappa, \omega)$ at $\bkappa = h \bk$,
square it and take the expectation value:
$P_F(h, \bk, \omega) \equiv  
\langle \bF(\bk, h\bk, \omega) \cdot  \bF^*(\bk, h\bk, \omega) \rangle$, then from 
(\ref{eq:powerspectrumdefinition}), (\ref{eq:Ffromtildef}) we have
\begineq
\label{eq:tildePestimate}
P_F(h, \bk, \omega) = {N_t^2 \over \tau} \sum\limits_{h'}
{d^2 k'\over (2 \pi)^2} P_{h'}(k', \omega - \bv_{h'} \cdot \bk')
W_x^2(\bk - \bk') W_\theta^2(h\bk - h' \bk') 
\endeq
where we have taken the integration time to be greater than
the eddy turnover time-scale to effect an integration over
spatial frequency.
As already discussed, a layer of turbulence has a spatio-temporal power spectrum which
is a flaring disk of thickness $\omega_*$ so the quantity 
$P_{h}(k, \omega - \bv_{h} \cdot \bk)$ appearing above is also a disk,
but is inclined with respect to the
$\omega = 0$ plane with mid-plane gradient $d \omega / d k = v$.
Furthermore, if the $D$-sized eddy turnover time is much less than the
translation time-scale $D/v$, as the observations indicate, 
then $\omega_* \ll v k_*$, so the 
vertical displacement of the
disk is large compared to its thickness. If $h$ coincides with an actual
deflecting layer, and if we consider for the moment only the contribution from
that layer $h' = h$, then $P_F(h, \bk, \omega)$ is a 2-dimensional convolution of 
this thin tilted disk with the function $W_x^2(\bk) W_\theta^2(h\bk)$.
Now this function has width 
$\delta k \sim {\rm min}(1/L, 1 / h \Theta)$ whereas the
intersection of the inclined disk with a plane $\omega =$ constant has
width $\Delta k \sim \omega_* / v$ so provided 
$\omega_* \gg v / {\rm max}(L, h \Theta)$
(i.e.~the eddy turnover time is short compared to the time-scale for the
eddy to be convected across either the entire field or the overall extent of the
array, whichever is greater) then $\Delta k \gg \delta k$ and
the convolution has little effect and therefore
\begineq
P_F(h, \bk, \omega) \simeq P_{h}(k, \omega - \bv_h \cdot \bk)
\endeq
and one can determine $P_{h}(\bk, \omega)$ by fitting for
inclined versions of disk models of the form (\ref{eq:pkwmodel}) directly to the
measured power spectrum $P_F(h, \bk, \omega)$
One particularly simple way to achieve this would be to compute
\begineq
P(h, \bv) \equiv \int d^2 k \int d \omega \; \Gamma(\bk, \omega - \bk \cdot \bv)
P_F(h , \bk, \omega)
\endeq
the local maxima of which should coincide with the heights, velocities of
the various layers.

\begin{figure}[htbp!]
\centering\epsfig{file=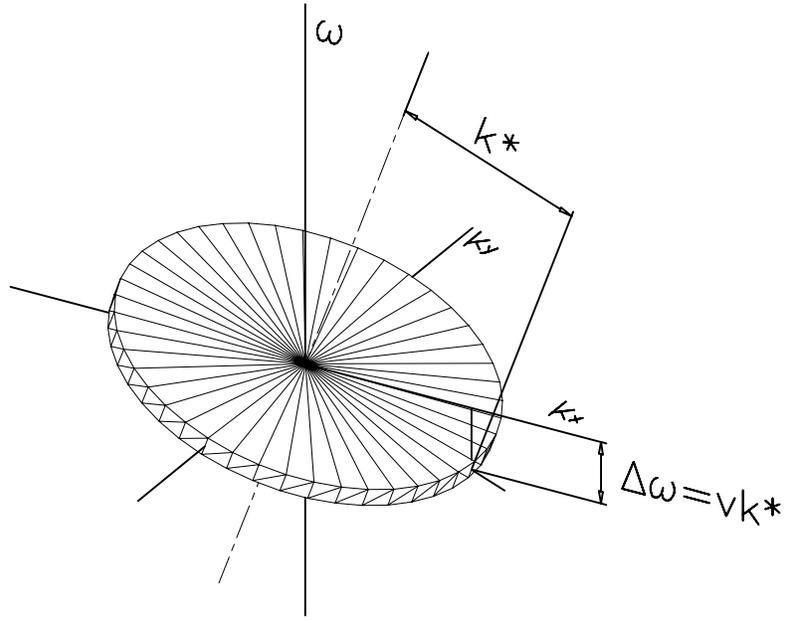,width=0.6\linewidth}
\caption[Spatio-temporal power spectrum for a streaming turbulent layer.]
{Schematic illustration of the spatio-temporal power spectrum 
for a streaming turbulent layer.  As discussed in the text, a
slowly evolving layer of turbulence has an intrinsic power spectrum $P_h(\bk, \omega)$
which is a disk-like function extending to $|k| \simeq 1/D$ in the
spatial direction and with thickness $\omega_* \sim 1 / t_D$
in the temporal direction (vertical in the figure).  The measured
power spectrum $P_F(h, \bk, \omega)$, with $h$ taken to be the height of some
layer streaming with velocity $\bv$, is
similar, but tilted slightly as illustrated, with the tip of the 
normal to the disk being displaced from the vertical axis in the direction
$\bv$.  Not shown in this figure is the power aliased from different
layers $h' \ne h$.  The aliased power consists of a periodic
array of weak `ghost' disks, tilted in the appropriate directions, and
with centers on a grid of points in the $k_x$, $k_y$ plane
with spacing $2 \pi / \Delta x$ where $\Delta x$ is the spacing
of the telescopes in the array.  
}
\label{fig:diskplot}
\end{figure}

It is important in what follows to have a clear picture of the form of the
power spectrum $P_F(h, \bk, \omega)$ which, being four dimensional, 
is somewhat difficult to visualize.  To this end, imagine
you are sitting in the control room of a wide field imaging
array.  Measurements of guide star deflections have been recorded and
transformed to produce $F(\bk, \bkappa, \omega)$ and you are
viewing a graphics device with a 3-D renderer
displaying an isodensity surface of the measured power $P_F(h, \bk, \omega)$ 
in $\bk$, $\omega$ space, integrated
over perhaps ten minutes or so, for a given altitude $h$.  
A widget on the screen allows you to control the altitude.
At first you see nothing.   As you slowly vary the altitude suddenly
a disk like structure springs into view, as shown schematically
in figure \ref{fig:diskplot}.
The radial extent of the disk has an Airy disk form as given by
the model (\ref{eq:pkwmodel}), and it has the expected bell-shaped
vertical profile.
Wiggling the altitude control you estimate that the
width of this structure is $\delta h \sim D / \Theta$. 
This is the signature of a horizontally convecting layer of seeing.
The disk is tilted (actually sheared), from which you can read off
the wind velocity vector, and the thickness tells you how fast
it is evolving internally.
Further variation of the altitude reveals a number of further
layers with different strengths and velocities and perhaps thicknesses,
but otherwise matching the pre-determined template.

Results of the kind described would lend credence to the idea that the
behaviour of the atmosphere can indeed be characterised by a tiny subset
of all the Fourier components computed, and that the
Fourier transform of the actual deflection at the present instant
${\bf f}_0(\bkappa)$ may be accurately given by a linear combination
of a small set of $\bF(\bk, \bkappa, \omega)$ values, and that this
might allow one to freeze out the motion of all the images in the field.
However, the display device also has a widget to control the level
of the isodensity contour plotted.  As you increase the contrast
the picture changes radically.  The disk centered on the origin
swells as expected, but rather suddenly a set of additional low level structures
appear laid out on a grid in the $\omega = 0$ plane.  
The spacing of these structures is $\delta k = 2 \pi / \Delta x$
where $\Delta x$ is the spacing of the telescopes --- so the
spacing is small compared to the extent of the disk ---
and on closer inspection they are seen to be well modeled by
superpositions of weak replicas of the disks seen in the high
iso-surface level scan, but with seemingly random strengths.  
This background of low level ghost images also persists when you tune the
altitude control to arbitrary positions.

What is happening here is aliasing of
deflection power from spatial frequencies differing from the
target frequency by integer multiples of $2 \pi / \Delta x$,
and from different heights $h' \ne h$.
In the foregoing we have assumed that 
$W_x$, $W_\theta$ are narrow functions of their arguments, and we
have approximated the values of various integrals by simply integrating
over the `central lobe' of these functions.  This is a good first approximation
to be sure, but in 
fact both of these functions have extended side-lobes.
For a regular grid telescope array, $W_x(\bk)$ has the
form of a 2-dimensional sinc function with central value
$W_x(0) = N_T^2$ and width $\delta k \sim 1/ L$ but this pattern repeats with
period $2 \pi / \Delta x$. 
Similarly, $W_\theta^2$ has a central lobe of
height $N_s^2$ and width $\delta \kappa \sim 1 / \Theta$, but
for $|\kappa| \gg 1/\Theta$ the function $W_\theta(\bkappa) = 
\sum e^{i \bkappa \cdot \btheta}$
is effectively the sum of $N_s$ random plane waves with random phases,
so it resembles a Gaussian random field with
coherence length $\delta \kappa \sim 1/\Theta$ and with mean square
value $W^2_\theta(\bkappa) = N_s$. 
These weak but extended side-lobes
will alias power from different spatial frequencies $\bk' \ne \bk$ and from
different layers $h' \ne h$ into $\bF(\bk, h \bk, \omega)$ but not into
${\bf f}_0(\bkappa)$. This leakage of power will result in imprecision in
the conditional mean estimator.  

To understand the conditions for obtaining an accurate deflection
model let us assume that we have correctly identified the 
the strengths and velocities of  the
deflecting layers, and that we have computed  $\bF(\bk, h \bk, \omega)$
for a spatial frequency $\bk \sim 1/ D$ and 
a temporal frequency lying within a particular layer; i.e.~for a point lying
within the disk shown in figure \ref{fig:diskplot}.  
The key
question is what fraction of $\bF(\bk, h \bk, \omega)$ actually arises
within the layer at the height $h$ and what fraction is aliased from
entirely different layers?  We can infer the answer to this question from
inspection of equation (\ref{eq:tildePestimate}).  This integral
will have a central lobe or `primary' contribution
and an integrated side-lobe or `aliased' contribution.
The primary contribution comes from $\bk' \simeq \bk$
and is $\sim P_h W_x^2(0) W_\theta^2(0) \delta k^2$ or
\begineq
P_F({\rm primary}) \simeq P_h N_s^2 N_T^2 \; {\rm min}(1/L^2, 1/h^2 \Theta^2)
\endeq
(we have ignored the prefactor $N_t^2 / \tau$).  The periodic form of
$W_x^2(\bk)$ results in a series of `ghost images' of the power for
all of the discrete layers, each with the appropriate tilt, and
replicated on a periodic grid in $\bk$ with spacing $\Delta k = 2 \pi / \Delta x$.
The aliased power 
therefore appears where $\bk - \bk' \simeq \bn \Delta k$, where $\bn$ is
a vector with integer values components.  These aliased contributions
are individually weak, because typically the argument of $W_\theta$ will be large compared to
$1 / \Theta$ so $W_\theta \simeq N_s$ rather than $N_s^2$.  
If there are $N_l$ layers then there will be
$\sim N_l \times (k / \Delta k)^2 = N_l (\Delta x / D)^2$ ghost images
located within the region of interest which is of size $k \sim 1/ D$,
so they are quite numerous. On the other hand,
the probability that a given point $\bk$, $\omega$ falls within
any one of these ghost images is $\sim D / {\rm max}(D, v t_D)$ which is
small if $v t_D \gg D$ which we expect to be the case.
Putting all these factors together we find for this integrated aliased
power
\begineq
P_F({\rm aliased}) \simeq P_h N_s N_T^2 (1 / L^2) \; {\rm max}(1, D / v t_D) 
(\Delta x / D)^2
\endeq
where we have assumed that
the velocities of the distinct layers are randomly distributed.  
Aliasing will be stronger from another layer which has nearly
the same velocity, but this is an unlikely situation.
The requirement that the primary contribution to $F(h \bk, \bk, \omega)$
(which correlates
strongly with the thing we are trying to predict) greatly exceed the
aliased contribution (which doesn't) is then simply
\begineq
\label{eq:aliasingcondition}
N_s N_T D \; {\rm max}(D, v t_D) \gg N_l\;  {\rm max}(L^2, h^2 \Theta^2).
\endeq
This is a key result of this section and is physically very reasonable.  
The RHS is the total area of all of the
deflecting layers, whereas the LHS is the 
total area of the samples of the deflection 
provided by $N_s$ guide stars seen through $N_T$ telescopes.  For aliasing to
be unimportant we need to sample the layers sufficently well.

Equation (\ref{eq:aliasingcondition}) provides the following important constraint
on the configuration of the array: one should not choose $L$ to be much greater than
$h \Theta$, which is about 100m for our canonical $h = 5$km and a 
$\Theta = 1$ degree field. Primarily this is because taking $L \gg h \Theta$
would unneccessarily increase the total area of deflection screen
that one must deal with.
This equation suggests that aliasing is independent of the array
size for smaller $L$.  However, this is not the whole story.
What we have neglected here is the possibility of aliasing
of power in (\ref{eq:tildePestimate}) from a 
nearby layer $h' = h + \Delta h$ through the 
central lobe of $W_x^2$.  Imagine we have taken $L$ to be very small so the
array is compact.
That means that the transform of the array $W_x^2$ will be correspondingly wide, 
with width $\delta k \sim 1 / L$, so in (\ref{eq:tildePestimate}) the factor
$W_x(\bk = \bk')$ will limit the contribution to the integral to a small region
of size $\sim 1/ L$ around $\bk' = \bk$. More restrictive however is the
factor $W_\theta^2(h\bk - h' \bk')$ which clearly has a narrow peak for $h' = h$
of width $\delta k \sim 1 / h \Theta$ around $\bk' = \bk$.  However, there will also be
secondary peaks of $W_\theta^2$ centered
on $\bk' = (h / h') \bk$, and these will fall within the more extended central lobe of
$W_x^2$ if $\Delta h / h \lsim \delta k / k \sim D / L$.  
This would be problematic: according to (\ref{eq:f0}) ${\bf f}_0(\bkappa)$
receives a contribution from each layer which is restricted
to a region of size $\delta k \sim 1 / h \Theta$ around $\bk = \bkappa / h$
whereas for $L \lsim D h / \delta h$, the measured $F(\bk, h \bk, \omega)$
receives comparable contributions from neighboring regions and this
will destroy the precision of our conditional mean estimate for ${\bf f}_0(\bkappa)$.
This is slightly over-pessimistic since 
the contribution from these secondary layers may in fact be
suppressed if their velocity difference is such that $\bk \cdot \Delta \bv > D \omega_*$.
This will {\sl usually\/} be the case if the internal evolution time-scale
is small compared to the convection time-scale, and if $\Delta \bv \sim \bv$,
but this cannot
be guaranteed, and in any case there will still be a range of $\bk$ values (lying
perpendicular to $\Delta \bv$)
for which the confusion cannot be resolved.  The solution to this problem is straightforward:
choose an array size $L$ which is satisfies $L \gsim D h / \Delta h$.
For a separation $\Delta h = 300$m say this gives $L \gsim 25$m.
  
There is a simple way to understand this latter constraint.  What we are doing
here is disentangling the deflections from distinct layers by evaluating
the 5-dimensional transform
$\bF(\bk, \bkappa, \omega)$ at the angular frequency $\bkappa = h \bk$
coresponding to a spatial frequency $\bk$ at height $h$.
If we use too small an array then we have poor resolution in spatial
frequency $\delta k \sim 1 / L$ and this converts to a corresponding
lack of resolution in height.  This harks back to the observation
made at the start of this section, that a layer of thickness $\delta h$
is only effectively thin if $L \lsim D h / \delta h$.
Thus, if the nature of the refractive index fluctuations is slowly
evolving (and therefore highly predictable)
streaming layers of thickness $\delta h$ with separation $\Delta h$ then
we want to be able to resolve the separate layers.  If we fail to
resolve two layers moving with different velocities, then we obtain
a single effective layer with enhanced width in temporal
frequency, which consequently has less
predictable evolution.  On the other hand, we do not want to
resolve the individual layers further into sub-layers.  If we
take $\delta h = 100$m and $\Delta h = 300$m then this would argue for
an array size $L \sim 50$m or so, this choice also being consistent with
the independent constrain that $L \lsim h \Theta$.

These considerations answer a question that may have occurred to the
reader.  Why not retro-fit an existing 10m class telescope with
some kind re-imaging optics with stops to generate
sub-pupils and with an array of cameras to synthesise
an array of 1.5m aperture telescopes?  While this approach is attractive
on grounds of cost, if the values $\delta h$, $\Delta h$
characterizing the stratification of seeing given above are 
at all accurate then such a design would be sub-optimal as 
compared to a purpose built array
with larger baselines because it would be unable to resolve the
separation between layers.
In the context of (\ref{eq:aliasingcondition}) such an instrument
would benefit from having a smaller effective number of layers $N_l$,
but the short effective coherence time for the `composite'
layers $t_D \sim D / v$ would tend to outweigh this gain.  

Returning to the case of an array, and assuming that the
condition $L \lsim h \Theta$ is satisfied, equation (\ref{eq:aliasingcondition})
becomes independent of the field size $\Theta$ since $N_s \propto \Theta^2$, and
gives then the number of telescopes required to deal with
a certain number of layers distributed over some range of heights.
As already emphasised, the sensitivity of the image quality to
imprecision of conditional deflection means that we want to
satisfy the inequality (\ref{eq:aliasingcondition}) strongly, but
even so we find equation (\ref{eq:aliasingcondition}) encouraging.
The arguments leading to (\ref{eq:aliasingcondition}) are admittedly hand-waving,
and it would be nice to quantify the dimensionless factors 
we have brushed over, but it is probably reasonable to scale
from the case of a single layer which we have analysed quantitatively.
As we saw in \S\ref{subsec:singlethinlayer}
for a high altitude ($h \sim 5$km) deflection layer 
a single layer is essentially fully constrained
with $\sim 10$ telescopes even in the absence of useful temporal
correlations.  So it should not be too difficult to satisfy the
condition (\ref{eq:aliasingcondition}) for realistic sized arrays.  
With say 30 telescopes one should be able to deal with several
high altitude layers and even more layers at low altitude.

With more detailed knowledge for the properties of the atmosphere it 
should be possible to accurately compute the performance of this
type of imaging system and predict the resulting image quality. 
This would then allow one to configure the telescope array to
optimize the performance. 
Unfortunately the data that are currently available from 
SCIDAR studies etc.~are sparse
and do not necessarily address the key issues here such as what is the 
time-scale for the layers to evolve etc.  What are critically needed
are measurements of the deflection correlation function which sample
the range of spatial, angular and temporal separations relevant here, and
also over a period of time in order to properly understand the
statistics of these meteorological processes.  
Some of these issues can
be addressed simply with drift scan observations on small telescopes,
or with measurements from wavefront sensors on
$8-10$m class telescopes,
but also needed are
measurements with two or more telescopes 
(or with a single large telescope equipped with pupil stops
and optics in order to simulate a number of small telescopes) in order to determine,
for example, the decorrelation at large angular separations due to finite
thickness of the deflecting layers.   
Experiments of this kind will be able to address the important, and to a large
degree open, question of intermittency and non-Gaussianity of 
the atmospheric deflections. Given data of this kind one can
then accurately establish the performance of the type of
instrument we are proposing before embarking on construction.

\section{Detectors}
\label{sec:detectors}

\def \arcsec {$''$}

The detectors for this array of telescopes must be able to compensate
for independent image motion on a scale of about 1 arc-minute, 
which for a field of view of about a degree
requires  3--4000 independent image motion compensation elements.
The other important consideration is that the
detector must have {\it many} pixels (approximately $10^9$ per square degree
for $\pixelsize$ pixels) for each of $\sim \Ntval$
telescopes, so the detectors must be reasonably inexpensive.

\begin{figure}[htbp!]
\centering\epsfig{file=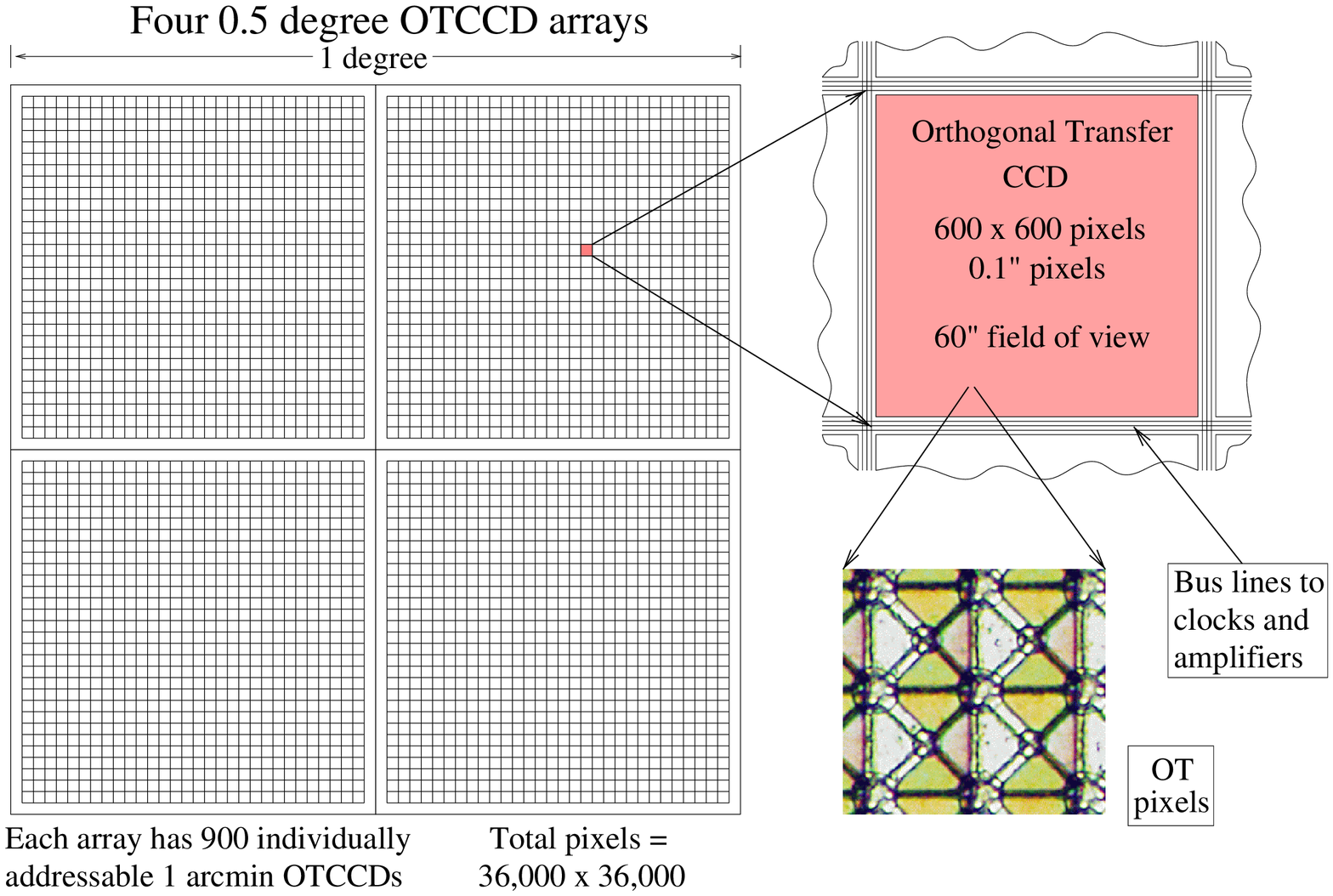,width=\figwidth}
\caption[Schematic of the proposed OTCCD array.]
{This schematic drawing illustrates the 1$^{\rm o}$$\times$1$^{\rm o}$ 
focal plane of independently addressable OTCCDs. The focal
plane consists of a 2$\times$2 array of 
$\chipsizeinpix \times \chipsizeinpix$
OTCCD arrays each measuring $\sim$100mm on a side. Each of
the small sub-cells measures 
$\cellsizeinpix\times\cellsizeinpix$ pixels and 
corresponds to a $\sim 1' \times 1'$ patch on the sky. 
The gaps between 
sub-cells are the equivalent of $\sim$30 pixels and are needed
to bus the various signals to the independent OTCCDs, and the larger
gaps between the monolithic arrays are needed for wire bond pads
and such.  
}
\label{fig:otarray}
\end{figure}

Our proposed solution to these challenges is a monolithic device
consisting of say a $\chipsizeincells\times\chipsizeincells$ array of
independently addressable $\cellsizeinpix \times \cellsizeinpix$ pixel
multi-directional or `orthogonal transfer' CCDs 
\cite{tbs97}.  With a pixel size of \pixsizeinmicrons $\mu$m the
resulting $\chipsizeinpix\times\chipsizeinpix$ pixel device would
measure $\sim$100mm$\times$100mm and would fill an entire 150mm diameter
silicon wafer.  With the optical system
designed to deliver a plate scale of
$\pixelsize/\pixsizeinmicrons\mu$m pixel, a $2\times 2$ array of
$\chipsizeinpix\times\chipsizeinpix$ OTCCDs butted together into a
mosaic would provide the desired 1$^{\rm o} \times 1^{\rm o}$ field
subdivided into independently-controllable 1$'$$\times$1$'$ patches.
Figure \ref{fig:otarray} illustrates this idea.  
The pixel size of $\pixsizeinmicrons\mu$m is a factor 3 times smaller than
those commonly used in astronomical applications, but similar to
the pixel sizes commonly used in consumer electronics applications.
Should manufacturing considerations force one to larger pixels one would
need to use a correspondingly larger array of chips to maintain
the desired field of view, with corresponding increase in costs.
The real limitations on physical pixel size is one of the major
unknowns in costing and optimizing this system.

In the subsequent sections we will describe how the OTCCD works and
how it can answer our need for a ``rubber focal plane'', the new step
of making a monolithic array of independent CCDs, how these arrays
would be operated in practice, and finally some estimates of
feasibility and yields.

\subsection{The Orthogonal Transfer CCD}

In any CCD the charge is localized into discrete pixels by application
of potentials to adjacent electrodes (``gates'') which are separated
from the charge collecting region by very thin layers of dielectric.
A CCD is read out by systematically changing the gate potentials in
such a way that the charge is moved laterally while maintaining each
pixel's identity by keeping at least one gate negative (potential
maximum) between each pixel.  A conventional 3-phase CCD achieves this
by using permanent implants (``channel stops'') which divide the CCD
into vertical columns and then three gates per pixel in the horizontal
direction.  This process of shifting the charge (``parallel
clocking'') is essentially noiseless (extra ``spurious'' charge
created by the process is typically unmeasurable for modest gate
voltages), highly efficient (charge transfer inefficiency (CTI)
is typically less than 1 part in $10^5$ or charge transfer
efficiency $\hbox{CTE} = 1 - \hbox{CTI} > 0.99999$), and
extremely fast (rates of $10^4 - 10^5$ pixel/sec are common).
There is no reason that this parallel clocking cannot take place while
the CCD is collecting light, and indeed this is the basis of time-delay
integration (TDI) or ``drift
scanning'' where a field of view is moved steadily down a column and the
CCD is read out at precisely the same rate so that the image is not
blurred (e.g. the SDSS CCD mosaic array).

\begin{figure}[htbp!]
\centering\epsfig{file=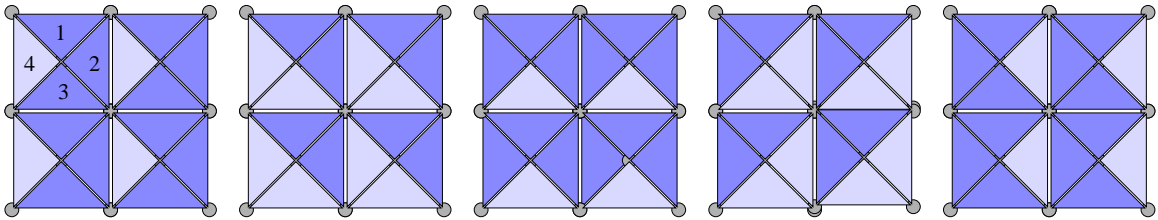,width=\figwidth}
\caption[OTCCD pixel gate structure and illustration of clocking.]
{This diagram shows the gate layout of a symmetrical OTCCD pixel and
illustrates how the charge is clocked to the right by
successive application of negative (dark) or positive (light)
potentials on the gates.  The gray circles represent
channel stops which prevent charge from moving between the corners of
the triangular gates.  The symmetry permits charge to be clocked
left, right, up, down, or even diagonally.  }
\label{fig:otccd_clock}
\end{figure}

The Orthogonal Transfer CCD (OTCCD) goes one step further in discarding
the permanent channel stop but introducing a fourth gate, and by making the layout
of the gates symmetric to $90^\circ$ rotations.  The OTCCD is
therefore capable of tracking image motion in an arbitrary direction.  As
the optical image dances about over the CCD, the accumulating charge
can be shifted in synchronism and any blurring from image motion will
be removed.  The device is bordered by a scupper which removes any
charge which is shifted off of the array.  Figure
\ref{fig:otccd_clock} shows how the charge is moved in such a device,
either for transport to the serial register for readout or for
tracking image motion, and the inset in figure \ref{fig:otarray} shows
a photomicrograph of the gates of an actual device with 15$\,\mu$m
pixels. 

\begin{figure}[htbp!]
\centering\epsfig{file=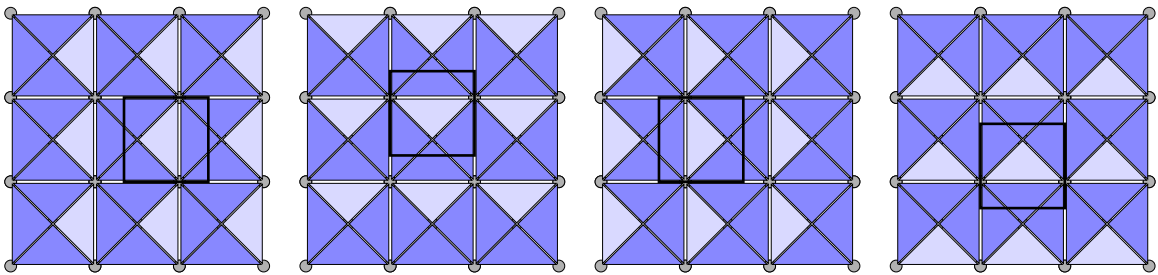,width=\figwidth}
\caption[OTCCD gate potentials for sub-pixel stepping.]
{By setting one of the four gates positive (light), the effective
location of a pixel can be shifted by a fraction of a pixel.  The heavy black
outline shows the region on the detector where photo-electrons will
migrate to a common point.
Other configurations using 2 or 3 positive gates are also possible,
allowing quite fine control of the collection pixel position.
}
\label{fig:otccd_substep}
\end{figure}

This layout of gates also lends itself well to fractional pixel
sub-stepping, which is important since we are expecting images with
$\FWHM$ FWHM and the pixel size is $\pixelsize$.
Figure \ref{fig:otccd_substep} shows how a collection pixel can be
shifted by a fraction of a pixel.  

\subsection{OTCCD Arrays}

Making a monolithic mosaic array of OTCCDs should not present serious
problems in design or manufacture.  The overall layout of the array
would have $\chipsizeincells\times\chipsizeincells$ independent
$\cellsizeinpix \times \cellsizeinpix$ OTCCDs separated by gaps
for common bus lines which might need to be as large as
about 30 pixels or 3\arcsec, which
is about a 10 percent effective dead area on the array.

The individually addressable OTCCDs would share common connections for
a number of signals such as serial gates and amplifier drain and
reset, but each must have independently addressable parallel gates (to
effect the necessary shifts for each CCD) and independent amplifier
outputs.  This can be accomplished by equipping each CCD with a set of
transistors which lie between these connections and the common bus
lines.  Using an x-y addressing scheme these transistors can be turned
on or off and any individual CCD connected or disconnected from the
external pins of the array.

As far as the external electronics is concerned, once 10 bits worth of
x-y address has been decoded and the appropriate CCD activated, the
array looks like a single, small CCD.

One challenge of making such an array work is that a CCD must maintain
the potentials on its gates or else it cannot keep the charge
within individual pixels.  However, the x-y multiplexing means that
all but one CCD is normally disconnected from the external drive
electronics.  We do not expect that this will be a problem, because
the capacitance of CCD gates is such that they will maintain their	
potentials for many seconds after being
disconnected.  Thus, as long as the CCDs are each periodically
reconnected to the external gate voltages they can continue to do
their job.  This mandates that the external electronics continuously
ripple through the array, much like dynamic RAM.  Since we need to
visit each CCD ten or more times a second to apply shifts, and
since we need many copies (perhaps 10 per array)
of the external electronics for speed in
acquiring guide star information, this should not be a problem.

A very significant advantage to this sort of CCD array is it is very
tolerant of manufacturing defects that would otherwise destroy a large
monolithic device.  Typical wafer defects are very localized, so a
defect that might render a conventional wafer-scale CCD unusable will
only ruin a single $\sim$1$'$$\times$1$'$ subarray on the OTCCD array.
Given that we are thinking of $\sim\Ntval$ telescopes each with its
own focal plane OTCCD arrays, the loss of a square arcminute from a
single detector is negligible.  Likewise the gaps between the CCDs
will not appear in the image produced by the entire telescope array,
since it is simple to offset the pointings of each telescope so that
the gaps do not overlap.

\subsection{Mosaic Operation}

Some number of the CCDs in this array will contain sufficiently bright
stars around which small patches will be read out for guiding
information, thus sacrificing the rest of that 1$'$$\times$1$'$ cell for
science.  This can be done in `shutterless video' mode and we have
already demonstrated that it is possible to work at 100~Hz
sampling with relatively slow electronics and relatively large CCDs
on available guide stars (see \S\ref{sec:guidestars}).

An example of the observational strategy follows: first, one would
identify the stars which are sufficiently bright to serve as guide
stars and identify the $\cellsizeinpix \times \cellsizeinpix$ pixel
OTCCD cells which contain these guide stars.  Then one would start the
following observing sequence. First all the CCDs are erased. Then a
3\arcsec$\times$3\arcsec sub-array surrounding each guide star is read
out from each OTCCD cell on each of the telescopes, and these are
pushed into $N_s \times N_T$ buffers of length $N_t$ which store
the most recent $N_t$ coordinates.

For each telescope, and for each cell of the detector, a deflection
vector is computed as a linear combination of the
data currently in the buffers (with matrix of coefficients computed on
the rather slow time-scale over which the deflection covariance matrix
evolves). The charge is then shifted in each of the OTCCD cells (those
integrating, and not those guiding) at each telescope to track the
motion.  A suitably filtered and averaged version of this guide signal
is fed to the telescope drives, so that the OTCCDs remove only the
rapid motion, and the guiding of the telescope keeps the overall
amplitude of OTCCD offset small.  Note that it is only this very
low frequency correction that is performed `closed-loop'.  The rapid
image motion correction is `open-loop', in the sense that there is
no feed-back from the guiding process on the guide stars themselves,
and this renders the high-frequency correction relatively stable.

Since this whole operation is completely parallelizable, 
the system can be run as fast as necessary by using 
enough drive electronics and
computers.  The communication between computers can easily be handled
by conventional Ethernet technology (with suitable attention to
latencies) since each telescope will have a local computer which
reduces all of the guide star images to a vector of $\sim 200$ offsets
every 50$\,$msec or so.

When the exposure is finished the shutter is closed and the arrays
read out.  Because we must necessarily have many read-out channels to
follow the guide stars, and since each $\cellsizeinpix \times
\cellsizeinpix$ CCD has its own amplifier, the readout can be
quite fast, possibly limited only by computer and memory bandwidth.
The CCD electronics should be able to read the entire array in under
30 seconds. An alternative is to keep the shutter open and
read out the integrating cells in some kind of ripple-through sequence
after relatively short (say 1-2 minute) exposures. Since the individual
cell read-out time is finite (about a second or so) this will result in
low level extensions of bright object images along the read-out
direction, but these can be removed
without difficulty.  Advantages of this approach are that 
it allows one to reject poor data if there are short bursts of bad
image quality, and it also allows greater time resolution for
transient events. 

\subsection{Feasibility and Yield}

To date OTCCDs have been produced in arrays as large
as 2K$\times$4K with 15$\,\mu$m pixels. 
Current lithographic techniques should allow the production
of $\pixsizeinmicrons\mu$m pixel devices which are similar in scale
to commercial consumer electronics applications, but 
whether this would have
significant impact on yield, quantum efficiency,
and blue MTF remains to be seen.   
One unavoidable effect will be a reduction in full well capacity to perhaps
20--50$\,$ke$^-$/pixel. This however is not a serious problem
when we consider that for this application 
the final telescope beam will be fairly slow, with 
focal ratio $f/8$ or thereabouts, and that the OTCCD cells
can be read out after quite short integration times.
Ideally, one would actually prefer even smaller pixels in order to
sample the PSF better, and that is a tradeoff which should be
explored.  

Despite the large size of such an OTCCD array, we believe the yield
for such devices would actually be relatively high.
We are already at the stage where manufacturers
are achieving extraordinary yields on large format
devices.  For example, it is not uncommon
for a lot of 150mm wafers filled with 2K$\times$4K CCDs to have
a yield exceeding 90\% and for more than 50\% of
these devices to be of scientific grade after thinning and
packaging (Burke, private communication). It now
appears to be feasible to fabricate a single wafer-scale device
with a high probability of success. Moreover, because of the
OTCCD array's high tolerance to defects, we expect that these
devices would have a much higher yield than a conventional CCD.

\section{Software}
\label{sec:software}

The software required by this project naturally breaks into two parts:
the software necessary to process the multiple guide star information
and carry out the fast guiding with the OTCCD detector arrays, and 
the software necessary to combine and process the resultant images.

The software required to operate the
CCD hardware is straightforward.  To the external world
the CCD array would have a single set of the usual gate and amplifier
connections as well as $\chipsizeincells + \chipsizeincells$ $x,y$ 
addressing lines which are set to 
access a given chip (we would probably include a 5-bit decoder for
each of $x$ and $y$ on the substrate).  Thus, once $x,y$ addresses are
set, the CCD array acts like no more than a conventional 
$\cellsizeinpix\times \cellsizeinpix$
pixel OTCCD.  This straightforward task could be carried out
by dedicated DSP-based CCD controller electronics, much like
those in used to operate current large CCD mosaics.

The software necessary to process the guide star information
and carry out the on-chip fast guiding at first
seems daunting, but in fact the massively parallel nature of this
process makes it much simpler than it might appear.  
For example, each telescope could 
have its control system interfaced to a master process which will ensure that
that telescope is pointing in the right direction and assemble
housekeeping information for the exposure.  This master process would 
also be responsible for shutters, filter wheels, and directing the
overall flow of guide star information from the CCD computers to the
covariance processes.
The rest of the software to run the OTCCD detectors is
straightforward and for the most part already exists. Routines 
have been developed to read out guide stars in shutterless
video mode, determine the centroid of the stars, and apply shifts to
OTCCDs.  The only new features that would be needed would involve
addressing individual cells to access the guide star information and
apply the OTCCD shifts.
However, this is precisely the same
code that is used for one CCD with a particular $x,y$ address enabled.
The natural loop that will be carried out in accessing guide star
information and applying shifts will also keep the gate voltages on the
individual CCDs refreshed.  If not for the necessity to combine all
the data from all the guide stars to compute the required shifts, 
the software task would be very simple.

The most challenging software task is the assembly of all the guide
star information into a covariance matrix.  Each of the CCD control
computers will be computing centroid information on perhaps 60
stars on a time scale of perhaps (50-100)~msec.  With 36 telescopes and 4
arrays per telescope, this is about 5000 sets of coordinate pairs
which are collected by two different processes.  The first process
uses an existing covariance matrix
and these new coordinate offsets to compute shifts for each
CCD in each array at each telescope.  These are shipped back to the
CCD control computers which apply them.  (Note that this does not have
to be synchronous -- it is better to ripple through the CCDs in
the arrays in a systematic manner.)  
The second process computes the covariance matrix.  Since this matrix
depends on things like the geometry of where the telescopes are sited,
the positions of the stars, and things like the upper atmosphere wind
direction and speed, or motions of eddies driving the outer scale of
turbulence, we expect that a given covariance matrix will be valid for
many minutes before its accuracy decays.  The time complexity of this
procedure was discussed in \S\ref{sec:correlations}.

Finally, the images must be combined and analyzed.
The data combination and analysis software
for the most part already exists.
The issues of CCD image processing,
image registration and combination, and PSF circularization are either
solved problems or will be by the time such a telescope array
might actually be built.
In many respects (e.g. good optical quality images and 
massive redundancy in the number of images being combined)
this telescope array eases many of the difficulties
typically encountered in CCD image combination.

\section{Overall System Cost Estimate}
\label{sec:costs}

Although the purpose of this paper is to outline a new
strategy for high-resolution wide field imaging, it
is nevertheless useful to estimate the cost of
such a system to investigate whether it is feasible
to build.
The system consists of $N_T$ telescopes, $N_T$ large
OTCCD mosaic cameras (1 per telescope),
a network of computers, the software to run the facility,
and some kind of building/enclosure.
Each of these is addressed below:

\subsection{Telescope Cost Estimate}
\label{sec:telescopecosts}

The modified RC telescope design with aspheric corrector
presents no significant problems. 
Similar sized units (with faster beams and wider fields)
have been constructed in recent
years (e.g.~USNO 1.3m telescope with a 1$^\circ$.7 field)
for \$0.7M or thereabouts. For this
array, each telescope, including the entire optical system, 
telescope mount, and telescope control system should cost 
under \$1M. The cost for a filled aperture, off-axis design may
be somewhat higher.  It is worth re-emphasizing here that the
cost of this system scales linearly with the collecting area, rather than
as some higher power as is the case for filled aperture large
telescope designs.

\subsection{OTCCD Mosaic Cost Estimate}
\label{sec:otccdcosts}

The OTCCD detector mosaic array is one area that presents 
a significant technical challenge and at first would appear to be
a very expensive development. However, as discussed earlier, the
architecture of the independently addressable array will
likely have very high yield because of its tolerance to
defects that would render a conventional large-format CCD useless. 
The estimate can therefore be based on the costs
to fabricate current, large 2K$\times$4K OTCCDs. Presently, it
takes approximately \$500K to fabricate a lot of devices on twelve
150~mm diameter wafers  and thin and package them. The large
OTCCD array would fill one wafer, so a lot would consist of
12 devices. It is perfectly reasonable to expect that at least
1/2 of the devices produced in such a lot will be usable
(the actual
number of good devices could be higher given our tolerance to defects
we mention above), therefore each $\chipsizeinpix\times\chipsizeinpix$
OTCCD array will cost less than or of order \$100K. 
This is in accord with the
idea that an OTCCD array should be of comparable cost to a single
2K$\times$4K OTCCD device because of the similar yields.
Four such detectors are needed for each camera to fill the
one square degree field, making the total cost for the detectors
alone of order \$400K/telescope. The additional costs per camera
are substantially lower than the detector costs. These
include the cost of the cryostat, the controller electronics, a
filter wheel, filters and a shutter.  All of these components are
similar to those for other large mosaic cameras currently in 
existence or under construction, and so
can be accurately costed.  The cryostat would employ a closed-cycle
cooler and can be readily built for under \$75K. 
The controller electronics would cost a similar amount 
($\sim$\$100K, assuming $\sim 30$
channels at \$2K apiece, clocking electronics, DSP controller, 
enclosure, etc.), and each camera
would require a filter wheel with filters and a shutter for about
\$50K bringing the additional hardware cost per camera to \$225K.  All
together, it is reasonable to expect that the total
detectors and camera hardware cost will not exceed \$625K/telescope, 
and thus is comparable to the cost of the telescopes themselves.
As previously mentioned, a big uncertainty here is the feasibility of
$\pixsizeinmicrons\mu$m pixels.  Should we need to use
say $7\mu$m pixels then the cost of the detectors would increase
by about \$500K/telescope

\subsection{Computer Network Cost Estimate}
\label{sec:computercosts}

The system would require a large network of parallel computers,
probably 1 per OTCCD array or 4 per telescope. Fortunately, these
computers would be extremely inexpensive as compared to the
rest of the hardware. The computer cost should not exceed
\$10K/telescope.

\subsection{Software Cost Estimate}
\label{sec:softwarecosts}

As expected, the hardware cost of the computers needed to
process the guide star information, carry out the fast guiding
on the OTCCDs, and combine and analyze the images, is
small compared to the software effort that will be needed. 
Nevertheless, the actual software tasks are manageable and
can be costed fairly accurately.
As discussed earlier, much of the software one can envision needing
already exists or will soon exist; image combination and analysis 
software can be adapted from the other data processing 
pipelines currently being designed and written. 
The most difficult task will be in handling the guide star information.
Not enough is known about
exactly what the atmosphere does to tip-tilt, particularly the
questions most pertinent here such as intermittency time scales, outer
scale sizes, stratification of turbulence, etc. 
Such an effort will likely involve both 
scientists as well as computer specialists,
and pilot project experiments will need to be carried out to learn
about how to combine multiple guide star information from multiple
apertures to compute image motion beyond the isokinetic angle.
It seems reasonable to expect that it will take some
30 man-years to develop the software needed to process
the guide star information, and another 5-10 man-years to
assemble the data analysis pipeline for the image combination
and analysis. Developing the DSP code to run the
OTCCDs will take approximately 1 man-year and other miscellaneous tasks 
might occupy another 1-2 man-years for a total of less than 43 man-years or 
approximately \$6.5M.

\subsection{Building/Enclosure Cost Estimate}
\label{sec:buildingcosts}

Unlike other large telescopes with rotating domes, the
building for this telescope array can be very simple. 
Depending on the spacing of the telescopes (for which
the optimal strategy needs to be worked out depending on 
the outcome of some experiments outlined in 
\S\ref{sec:correlations} above) there are several 
options for the enclosure: 1) if the telescopes are arranged
in a close-packed array, they could be mounted on some kind of
elevated frame with a roof and sidings which roll back to
allow the low-level boundary layer fluctuations to pass
safely underneath.
2) if the telescopes are
widely spaced, then each telescope might have its own 
small enclosure; or 3)  the telescopes might be grouped in several 
clusters, each with its own building. In all cases, the building(s) can
be relatively simple, and
should be substantially lower cost that the typical 8m telescope dome.
A reasonable upper limit would seem to be \$5M for the
enclosure.

\subsection{Total Cost Estimate}
\label{sec:totalcosts}

The collecting area of the telescope array scales as the number of
telescopes $N_T$.
Ignoring the enclosure and software, the cost of this telescope 
array also scales as $N_T$. 
Ideally one would determine the cost and performance as a function of
$N_T$ and other parameters such as the number of chips per camera
and then solve for the optimal performance and price.
Unfortunately however, determining the performance as a function of $N_T$ say
requires knowledge of the atmospheric conditions which is not yet
available.  In the absence of reliable information we will
assume that $N_T = \Ntval$ telescopes is adequate 
for which the total cost would be $\sim \totalcost$ and would yield 
the equivalent of a 9m aperture 
telescope that can deliver $\FWHM$ FWHM images over a 
1$^{\rm o}$$\times$1$^{\rm o}$ field of view. 
For $7 \mu$m pixels the cost would be larger by about $\sim \$20$M, as
would also be the case for detectors with $5 \mu$m pixels but twice the
field of view. 
One huge advantage of
this approach is that the aperture can grow to arbitrarily large
size simply by adding more telescopes. Conversely, one can start
building such an array and start using it once only a fraction of the
full complement of telescopes
are in place. Science observations can begin before the
final configuration is completed.

\section{Science}
\label{sec:science}

\def \arcmin {'}

The original motivation for the
wide-field high resolution imager (WFHRI) that we have described here
was to give enhanced resolution images for weak lensing.
However, it is potentially an exceedingly powerful instrument for many 
other applications.
It can be used in two different modes: the
``High Resolution Mode'' described here where all telescopes point in
the same direction in order to improve the PSF, and a ``Wide Field
Mode'' where the telescopes are pointed in different directions and
accept the natural seeing at the site.  Note that in Wide Field Mode
one can observe 36 square degrees simultaneously with a 1.5-m aperture.

\subsection{Performance of the WFHRI for Weak Lensing Observations}
\label{subsec:lensing}

A major driver for this imaging system was to provide
enhanced performance for weak lensing observations, which are
particularly hampered by poor resolution.  
In High Resolution Mode we expect to obtain FWHM $\FWHM$ images much of
the time over a 1 degree field, with an effective 9-m aperture.  
This is {\it much} more effective for measuring weak
lensing than a conventional large single aperture telescope
images which of much poorer resolution.  The net
result is that we will measure more galaxies, which gives us more
spatial resolution and precision for measuring weak shear, and as
we will now show, the improvement in
sharpness of the PSF results in an increase in performance for
shear measurements from small faint galaxies of better than
a factor 100.

It is possible to
define a quantitative figure of merit, the
inverse shear variance per unit solid angle, which measures the
power of given image data for weak lensing shear measurements
\cite{kaiser99}.   A detailed analysis is beyond the scope of this
paper, but the following simple argument should give a reasonable indication
of the increase in shear precision that will be allowed by
fast guiding.

Consider a population of objects, which we will model 
as small Gaussian ellipsoids with semi-major axes $a, b$, so the
intrinsic brightness distribution is $f(\br) = \exp(-0.5(x^2 / a^2 + y^2 / b^2))$.
Now model the PSF as either a single Gaussian for uncorrected
imaging, or as a double Gaussian for fast guiding.  In the latter case, and
for small objects which we know from e.g.~the HDF dominate the faint galaxy
population, essentially all the information will be contained in the 
core component, so we can equally compare the performance of two single
Gaussian PSFs, where the fast guiding PSF has a smaller scale length $\sigma$
but contains
only a fraction $f$ of the light.  After convolving
with the PSF, the object will be a Gaussian ellipsoid with semi-major
axes $A, B$ where $A^2 = a^2 + \sigma^2$, $B^2 = b^2 + \sigma^2$.
A shear $\gamma$ gives a net asymmetry for galaxies $\langle a^2 - b^2 \rangle
\simeq \gamma \langle a^2 + b^2 \rangle$.  A simple shear estimator is then 
$\hat \gamma = \langle A^2 - B^2 \rangle / \langle A^2 + B^2 - 2 \sigma^2 \rangle$.
When we average over a large number of similar galaxies the
denominator converges to $\langle a^2 + b^2 \rangle$, the 
average intrinsic area of the objects and, for small shear at least,
the uncertainty in the shear estimator is
dominated by fluctuations in the numerator.  

The fractional measurement error in $A, B$ is
$\Delta A / A \sim \sqrt{N_{\rm bg}} / N_{\rm obj}$ where
$N_{\rm bg}$ is the number of photons from the sky over the 
object, assumed to dominate over the count of photons from the
galaxy, and is  proportional to $A^2$ and $N_{\rm obj}$ is the number of
photons detected from the object, which is proportional to
$f$, hence as far as the dependence on PSF properties is concerned
we expect $\Delta A / A \propto A / f$ (we are assuming that the
measurement error is of similar order to the intrinsic noise due to
the inherent shapes of galaxies).  The uncertainty in the
shear estimator is then $\Delta \hat \gamma \sim A^2 \Delta A / A
\propto A^3 / f$ and the performance of the instrument $P$ is
proportional to the inverse square of the shear error, that is
$P\propto f^2 / A^6$. For poorly resolved objects the post
convolution size is $A \simeq \sigma$, so the performance therefore scales
as $P \propto f^2 / \sigma^6$.  

For the pixellated peak-tracking PSF (with $r_0 = 40$cm and
$\lambda = 0.8\mu$m) we find that the core has
$\sigma = 0''.05$ and contains about $f = 28$\% of the light
whereas a single Gaussian fit to the uncorrected PSF gives
$\sigma = 0''.17$ so we find that 
$P = (0.28)^2 (0.05 / 0.17)^{-6} \simeq 120$.
Thus, holding all other factors such as net collecting area,
detector efficiency etc.~constant, this says that sharper shape of the
PSF as compared to the uncompensated case yields better than
two orders of magnitude improvement in efficiency.  Equivalently,
to achieve the same precision in shear measurement from small, faint
galaxies one would need to integrate on the order of
100 times longer with a conventional telescope than with the
system we are proposing here.

\subsection{Other Science}
\label{subsec:otherscience}

High Resolution Mode is also valuable for many other projects.  For
example, with FWHM $\FWHM$ images at $\lambda \sim 1 \mu$m, we can measure
distances of galaxies out to about 10,000~km/s with 5\% accuracy using
surface brightness fluctuations.  The 1 degree field of view means
that we can observe much of an entire galaxy cluster at once,
obtaining hundreds of distances, and the 9-m aperture will give us
sufficient photons for this measurement at 10,000~km/s with an
exposure time of about 20 minutes.  Clearly the WFHRI has incredible
potential for mapping out the large scale flows in our local universe.

Another project for High Resolution Mode is simply to select a
non-trivial piece of sky and take very deep images in multiple
colors.  The value of the Hubble Deep Fields for e.g.~studying internal
structure in young, high redshift galaxies
cannot be underestimated,
but they were very expensive to obtain, and it is clear from the
differences between the northern and southern fields that a
$2'.5$ field is too small compared to cosmological structures.
The WFHRI compares favorably with HST for this project.  The
imaging performance will be slightly worse, due to the slightly
smaller aperture, although the pixel sizes
are comparable.  However, the WFHRI is a factor of about 50 more
sensitive than HST and WFPC2 in collecting photons, and it has a field
of view which is 500 times larger.  Given that much of the science
from the Hubble Deep Fields is not compromised by a factor of two
worse resolution, the WFHRI is arguably a factor of 25,000 more
efficient than HST, while being considerably less expensive.

There are many other projects which could be tackled with the WFHRI,
for example micro-lensing in the galactic bulge or M31 in order to
detect MACHOs.  The essential figure of merit in these studies is the
number of stars you can monitor.  The WFHRI has a huge advantage over
any other telescope existing or planned because of its large aperture
and superior PSF.  Another project for which the WFHRI is very well
adapted is searching for high redshift supernovae.  At redshifts
greater than about 0.5 (and even more so for $z > 1$) a 1 degree field
of view is more than adequate, but discovery and followup is {\it
vastly} easier with FWHM $\FWHM$ imaging than with $0''.5$.

In Wide Field Mode, one can point the 36 telescopes in a square array
covering 36 square degrees, and accept the 0.4--0.5\arcsec\ seeing
which results from being restricted to high-speed tip-tilt compensation
over an entire array. 
The array can also be operated in this mode whenever high altitude
turbulence is relatively weak, and the image quality would then
be similar to high resolution mode.
In this mode the WFHRI could survey the entire
northern sky in only $\sim 500$ pointings.  The sensitivity of a 1.5-m
aperture in the $R$ band with 0.5\arcsec\ imaging and an exposure time
of 300 sec is $R\sim24$ for a 5-$\sigma$ detection.  Given that one
could expect to get $\sim 100$ such observations per night, the WFHRI
could survey the visible sky (20,000 square degrees) in 5 nights.

In Wide Field Mode the WFHRI would be extremely useful for searching
for Kuiper Belt objects, both faint and bright, and for searching for
near Earth objects (NEOs).  As is true with the Dark Matter Telescope, 
following a systematic survey scheme means that finding and obtaining
orbits for asteroids and comets is done automatically by the fact that
the entire sky is surveyed repeatedly. 

A useful figure of merit for survey telescopes is
$M = A\Omega\eta/\Omega_{\rm PSF}$, where $A$ is the collecting
aperture, $\Omega$ is the survey area, $\eta$ is the detector quantum
efficiency, and $\Omega_{\rm PSF}$ is the solid angle of the point
spread function.  The difficulty, as we have seen, is that there
is no unique way to characterize the PSF area; the gain in performance
depends strongly on the application.  The faint galaxy application for
which the enhanced resolution is {\sl least\/} useful is 
simply detection, for the purposes of making counts of galaxies and
performing angular correlation studies etc.  It is interesting
to compare the performance of the WFHRI for this task with
other specialized proposals for wide-field survey instruments.
A ``Dark Matter Telescope'' has recently been proposed by
Tyson and Angel \cite{dmtelescopewebsite} which would have a 6.9m
effective aperture, a $3^\circ$ diameter field of view (7 square degrees)
and which would be expected to provide a typical PSF of $0''.6$ FWHM.
In Wide Field Mode, the WFHRI covers 5 times as much field of view as
the DMT ($6^\circ\times6^\circ$ versus $3^\circ$ diameter), has 1/20
the collecting area (1.5-m aperture versus 6.9-m effective aperture),
has similar quantum efficiency detectors, and will have distinctly
better imaging performance due to the smaller apertures and the
ability to do fast tip-tilt compensation for the entire array
(amounting to an improvement of perhaps a factor of 0.7 in PSF size
for given atmospheric conditions).

In High Resolution Mode, the WFHRI covers 1/7 times as much field of view as
the DMT ($1^\circ\times1^\circ$ versus $3^\circ$ diameter), has 1.7
the collecting area (9-m aperture versus 6.9-m effective aperture),
has similar quantum efficiency detectors, and will have imaging
performance which will be better by a factor of 3
(0.2\arcsec\ versus 0.6\arcsec).
Putting these together for the DMT yields 260~(m-deg)$^2$, taking
$\eta$ and $\Omega_{\rm PSF}$ to be unity.  In Wide Field Mode,
the figure of merit is 130~(m-deg)$^2$ for the WFHRI (applying the
factor of 0.7 to the PSF size), and in High Resolution Mode, the
figure of merit is 400~(m-deg)$^2$ (applying the factor of
1/3 to the PSF size).  Thus, for shallow, very wide field surveys the
WFHRI is slower than the DMT by a factor of 2 (although the DMT may
have trouble achieving a high duty factor because of detector readout
and telescope slew time), and for deeper surveys the WFHRI is faster
than the DMT by a factor of 2.  These numbers should not be considered
definitive since the optimal configuration for the WFHRI is as yet
not well known.  In particular, it is quite feasible to increase the
field of view substantially, though with some increase in detector costs.

\begin{figure}[htbp!]
\centering\epsfig{file=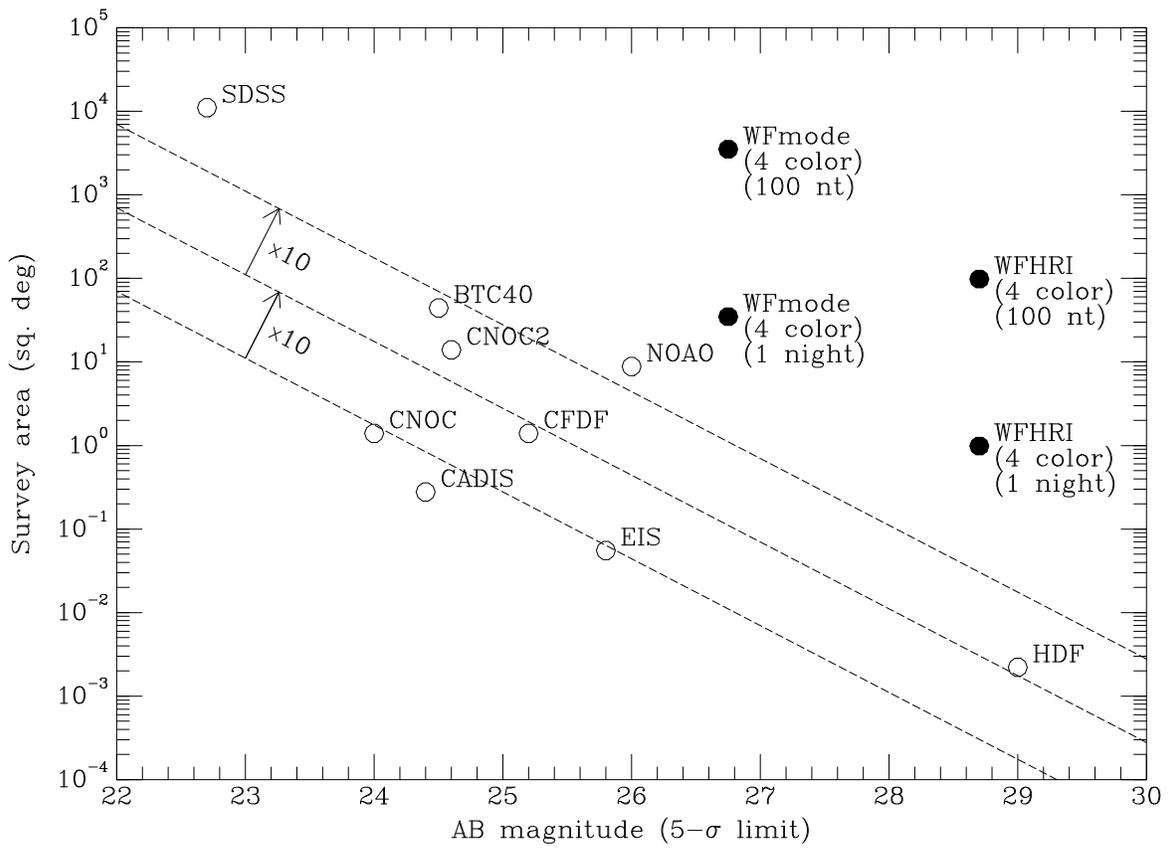,width=\figwidth}
\caption[Relative performance of various surveys.]
{Relative performance of various surveys.}
\label{fig:surveys}
\end{figure}

Both the DMT and the WFHRI are orders of magnitude better survey
telescopes than anything existing or currently planned.  As figure 
\ref{fig:surveys}
illustrates, the WFHRI can achieve the depth of the HDF in one night,
except over a square degree and in $B$, $V$, $R$, and $I$ (each to an AB
magnitude of 28.7).  Alternatively, in wide field mode the WFHRI can
rival the breadth of the SDSS, except going four magnitudes deeper.

We believe therefore that the WFHRI is a superior concept because it
offers much better performance for key scientific projects (and is
applicable to many additional projects), it is a faster
survey telescope for many applications such as searching for very
faint rare objects, and it is extremely flexible in the way it can be
deployed for a particular scientific objective.

\section{Conclusion}
\label{sec:conclusion}

We have outlined a strategy for wide field imaging at optical wavelengths with high
angular resolution by means of low-order AO in the form of fast guiding. 
Any AO system for wide-field applications must address
the `isoplanatic angle' problem --- that different parts of a wide field
will suffer largely independent wavefront distortions --- and so
it is necessary to multiplex the wavefront correction process.  
The solution here is limited in that we only attempt to correct for
the very lowest order wavefront distortions (though we do this
separately for a large number of small telescopes) but exploit the new
technology of OTCCD devices to efficiently apply fast guiding
independently to each of a huge number of isoplanatic patches.
The key feature of this strategy is the
ability to provide a PSF where roughly 30\% of the light is in a
very  small, diffraction limited core with FWHM of about $\FWHM$.
This is modest resolution compared to full AO on a large telescope, but
nonetheless would provide great gains in performance for many
applications within the general context of wide field imaging.   

In this paper
we have made detailed analytic and numerical calculations of the
expected image quality and we have quantified the constraints
implied by the limited numbers of potential guide stars.
We have computed in some detail how image quality can be improved for
the simple case of a single deflecting layer and we have described one
possible approach for extending this to the multi-layer case.
In particular, we have tried to set out the general constraints
that a system of this kind must satisfy in order to give the full
improvement in image quality.  
In common with multi-conjugate AO systems, a
key assumption here is that the source of seeing is highly
stratified; the type of system we have proposed
would be impractical if the source of atmospheric seeing were
distributed quasi-uniformly with altitude.  
Thankfully  strong stratification is
indicated by the currently available results of site testing using
SCIDAR etc., but more information is needed.  

We have argued that
a modest program to measure deflection correlations over the
range of angular, spatial and temporal separations relevant here
would allow one to definitively establish the performance
of this kind of instrument before constructing the full scale instrument.
Once these meteorological parameters are better determined
the next logical step would be to make a detailed
cost {\sl vs\/} performance analysis to compare this approach
with e.g.~a specialized wide field survey telescope in space.

We have described the design and operation
of orthogonal transfer CCDs; how these can be combined in large
numbers in single wafer-scale devices with high yield,  
and we have outlined the operating
procedure for contolling and collecting data from such a mosaic camera.
We have given what we feel to be a fairly conservative estimate of the
various component costs for a WFHRI.

We have tried to quantify the gain in performance for various
faint object applications as compared to
conventional terrestrial observing. Crudely speaking
these fall into two categories.  For simple detection and counting
studies, the gain is the least with a factor 2-3 improvement from image
quality.
The relatively modest gain in efficiency in the face of a fairly large
decrease in FWHM is not too surprising when one recognizes that the
effective area for collecting target object photons into the PSF core
is a fraction of the actual collecting area, but the full area is
effective for collecting sky background photons.  For sky
noise dominated photometry the performance is poorer than for
a hypothetical telescope with $1/3$ the collecting area but
which can concentrate 100\% of the photons in a similarly compact PSF.
The applications for which the gain is the greatest ---
up to about 2 orders of magnitude gain from image quality alone ---
are those that really demand high resolution.  Such applications
are those that need the spatial frequencies in the image which are exponentially
suppressed in uncorrected imaging by a huge factor but are preserved in the
WFHRI at levels only a few times lower than for a diffraction
limited telescope.  Examples that require this very high resolution
are typically those which explore the {\sl structure\/} of galaxies,
such as weak lensing and studies of star formation and related
morphological evolution in faint galaxies, or 
those applications which require highly
accurate positional data.

One of the real strengths of the WFHRI is its flexibility.  We have
described high resolution and wide field modes, but of course it is
possible to use a WFHRI in other modes as well.  For example, under
favorable conditions and with say
clusters of six telescopes, WFHRI can survey 6 square degrees
simultaneously with an effective aperture of 3.6~m, and nearly full
improvement in Strehl ratio.  Also, as we have stressed, the WFHRI is not
an ``all or nothing'' proposition.  For not much more than 1/6 the
cost, a working array of six telescopes could be put into operation 
which would be comparable to the Megaprime imager being built for the
CFHT, except that the PSF would be a factor of 2--3 better.
Before new large telescopes are built for wide field imaging, 
this new alternative approach deserves serious consideration.

We would like to thank Steve Ridgway for constructive criticism of
an early draft of this paper. Buzz Graves explored several optical designs for
wide-field telescopes that convinced us that such systems are straighforward to 
build with current technology.
We also gratefully
acknowledge many helpful discussions with Malcolm
Northcott and helpful comments from Chris Stubbs.

\end{document}